\documentclass[longbibliography,
preprint,
 amsmath,amssymb,
pra,
nobalancelastpage,
floatfix,
twoside,
]{revtex4-1}

\usepackage{balance}

\usepackage{flushend}

\usepackage{fancyhdr}

\usepackage{soul}
\usepackage{color}
 \usepackage[dvipsnames]{xcolor}
\newcommand{\ze}[1]{#1}
\newcommand{\rework}[1]{#1}
\newcommand{\sework}[1]{#1}

\newcommand{\me}[1]{#1}
\newcommand{\jpapT}[1]{#1}
{
  \fancyhf{}
  \fancyhead[LO]{\MakeUppercase{Non-equilibrium Green's Function and First ...}}
  \fancyhead[LE]{\MakeUppercase{Andrawis} and ROY}
  \fancyhead[RE]{}
  \fancyfoot[C]{\thepage}%

}
\fancypagestyle{firstpage}{%
  \lhead{}
  \rhead{}
}
\RequirePackage{lineno}
\usepackage[bottom]{footmisc}

\pdfoutput=1
\usepackage{subfigure}
\usepackage[colorlinks=true,linkcolor=black,citecolor=black,urlcolor=black,bookmarks=true,hypertexnames=true]{hyperref}
\usepackage{graphicx}
\usepackage{dcolumn}
\usepackage{bm}

\begin{document}

\raggedbottom
\preprint{APS/123-QED}

\title{Non-equilibrium Green's Function and First Principle Approach to Modeling of Multiferroic Tunnel Junctions}

\author{Robert Andrawis}
\email{randrawi@purdue.edu}
\author{Kaushik Roy}%
\affiliation{%
  School of Electrical and Computer Engineering,
Purdue University, West Lafayette, Indiana 47906, USA\\
(Submitted to peer review in 9 November 2018)
}%
\begin{abstract}
Recently, multiferroic tunnel junctions (MFTJs) have gained \sework{significant} spotlight in the literature due to its high tunneling electro-resistance together with its non-volatility. In order to analyze such devices and to have insightful understanding of its characteristics, there is a need for developing a multi-physics modeling and simulation framework. The simulation framework discussed in this paper is motivated by the scarcity of such multi-physics studies in the literature. In this study, a theoretical analysis of MFTJs is demonstrated using \me{self-consistent analysis} of spin-based non-equilibrium Green's function (NEGF) method to estimate the tunneling current, Landau-Khalatnikov (LK) equation to model the ferroelectric polarization \sework{dynamics}, together with landau-Lifshitz-Gilbert's (LLG) equations to capture the magnetization dynamics. The spin-based NEGF method is equipped with a magnetization dependent Hamiltonian that eases the modeling of the tunneling electro-resistance (TER), tunneling magneto-resistance (TMR), and the magnetoelectric effect (ME) in MFTJs.  Moreover, we apply the first principle calculations to estimate the screening lengths of the MFTJ electrodes that are \sework{necessary} for \ze{estimation of tunneling current}. The \sework{simulation results of the proposed framework} are in good agreement with the experimental results. Finally, a comprehensive analysis of TER and TMR of MFTJs and their dependence on various device parameters is illustrated.
\end{abstract}
\maketitle
\thispagestyle{firstpage}
\section{Introduction}

Over the last few decades, the complementary metal-oxide-semiconductor (CMOS) technology has been continuously downscaled \rework{following}  Moore's law \cite{Moore1998}. However, the static power dissipation and the threshold voltage variations of downscaled short channel transistor \rework{have} become dominating factors that limit the \rework{static random access memory} (SRAM) performance \cite{Chang2013TechnologyEDRAMb, Ye2010RandomCircuits, Chun2013AMemoryb}. \rework{Consequently, the high static power dissipation of SRAM inspired the exploration of alternative memory technologies like spin transfer torque magnetic memory (STT-MRAM).} However, the limited tunneling magneto-resistance (TMR) of magnetic tunnel junction (MTJ) together with the threshold voltage fluctuations of the short channel access transistor \rework{affect} the STT-MRAM read error rate. Therefore, the read performance of STT-MRAM \ze{has become} a fundamental limiting factor \sework{in} its applicability. Consequently, a new family of tunnel junctions, called ferroelectric tunnel junctions (FTJs), have emerged in literature \cite{Chanthbouala2012Solid-state, Garcia2014, Yin2017}.

\begin{figure}[!ht]
\centering
\includegraphics[width=1\linewidth,trim={0cm 0.7cm 0cm 0cm}]{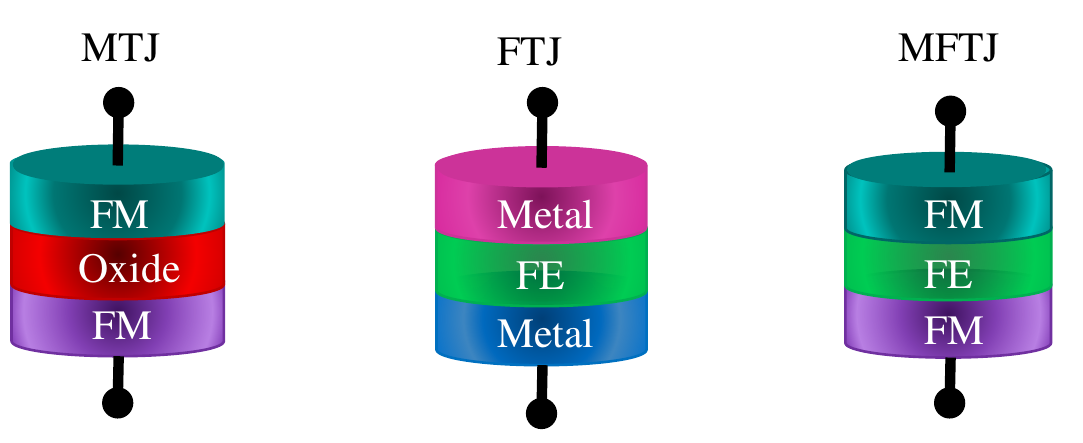}
\caption{The layer structure of MTJ, FTJ, and MFTJ.}
\label{fig_MFTJ}
\end{figure}

\rework{An} FTJ consists of a ferroelectric insulator sandwiched \rework{between} two different metal electrodes, as illustrated in Fig. \ref{fig_MFTJ}. The information is stored in the electric polarization of the insulator. The FTJ resistance is a function of the electric polarization of the insulator. The electric polarization of the ferroelectric insulator modulates the FTJ resistance, and hence the information \rework{can be} extracted by sensing the FTJ resistance. The tunneling electro-resistance (TER) of FTJ is defined as $TER=\frac{|R_\rightarrow-R_\leftarrow|}{min(R_\rightarrow,R_\leftarrow)}$, where $R_\rightarrow$ and $R_\leftarrow$ are the resistance of positive and negative electric polarization states, respectively \cite{esaki1971polar}. The physical origin of TER is discussed in detail in section \ref{ch_formalism}. The charge current of FTJ consists of three main components: Fowler–Nordheim tunneling, direct tunneling, and thermionic emission \cite{Pantel2010}.

On the other hand, the multiferroic tunnel junctions (MFTJ) is a nonvolatile tunnel junction that consists of two ferromagnetic layers separated by a ferroelectric insulator, as illustrated in Fig. \ref{fig_MFTJ}. Intuitively, from the structure of an MFTJ, we can predict that an MFTJ combines the resistive switching mechanism of FTJ and MTJ to constitute a four-state device. However, it \rework{turns} out that the MFTJ has more advantages over its constituent devices due to the magnetoelectric effect (ME). The ME effect at the FM/FE interface is observed in LaSrMnO$_3$(LSMO)/LaCaMnO$_3$(LCMO)/BaTiO$_3$ (BTO)/LSMO MFTJ \cite{Pantel2010}. \sework{\ze{It} originates from the modulation of the screening charges at the LCMO side by the bound charges at the BTO interface. The change in the \ze{electron} concentration at the LCMO interface affects the LCMO magnetic configuration. The magnetic alignment of the LCMO layer is switched from the ferromagnetic (FM) to the antiferromagnetic (AFM) alignment due to the change in \ze{electron} concentration \cite{burton2009prediction, Yin2015}.} \rework{The transition to the AFM alignment shifts the density of states (DOS) of the majority spin carriers to higher energy levels, and hence limits the majority spin current. In brief, the overall influence of the ME effect is to improve the TER ratio, as explained in detail in section \ref{ch_formalism} and section \ref{ch_dft}.}

A detailed review of the state of the art in FTJs and MFTJs could be found in \cite{Garcia2014, Yin2017}. However, a brief review of the progress in FTJs and MFTJs literature is provided in the following discussion.  Although the FTJ has been predicted by Esaki \textit{et al.} \cite{esaki1971polar} in 1971 with the name "polar switch", the FTJ has not been realized until recently. The lack of the knowledge of fabrication techniques of ferroelectric ultra-thin films had prevented the FTJ realization. However, due to the breakthrough that has been achieved by Zembilgotov \textit{et al.} \cite{Zembilgotov2002Ultrathin}, the ferroelectric ultra-thin film has been realized followed by many other experimental studies \cite{Hongwei2003Nanoscale, RodrguezContreras2003}. Zhuravlev \textit{et al.} \cite{Zhuravlev2005, Zhuravlev2009} have explained the dependence of the barrier height on electric polarization with the help of Thomas-Fermi equation and Landau tunneling current formula \cite{Appelbaum1244}. The Wenzel-Kramer-Brillouin (WKB) approximation and one-band model \cite{Appelbaum1244,burstein1969tunneling,wolf1985principles} have been used to calculate the tunneling current through the FE insulator in \cite{Kohlstedt2005}. Hinsche \textit{et al.} have used Landauer-Büttiker formula and the WKB approximation together with \textit{ab initio} calculation to model the FTJ characteristics \cite{hinsche2010strong}. Fechner \textit{et al.} have used the \textit{ab initio} method to study the electric polarization dependent phase transition in Fe/ATiO$_3$ interface \cite{fechner2008magnetic}. On the other hand, the non-equilibrium Green's function (NEGF) method along with Landau-Khalatnikov (LK) equation have been used to estimate the FTJ I-V characteristics in \cite{Chang2017}. However, the study did not consider the magnetization dynamics or the Hamiltonian dependence on the magnetization.

To conclude, the scarcity of multi-physics simulation studies that capture the MFTJ magnetization dynamics, \sework{along with} TMR, and TER effects motivates the \sework{modeling and} simulation framework applied in this study. In this study, the spin-based NEGF is applied to model the tunneling current, along with landau-Lifshitz-Gilbert's (LLG) equation are applied to model the magnetization dynamics, and the LK equation is applied to describe the FE motion. However, the accuracy of these models depends on the parameters used to model various materials. \ze{In} \sework{our} study, we use the density functional theory (DFT) \sework{to} estimate the electrostatic potential, and hence the screening lengths of the electrode that are used in the NEGF transport simulations. \ze{The} simulation results are compared to experimental results of the MFTJ in \cite{Yin2015, Chanthbouala2012Solid_state} to confirm the validity of the method.

\jpapT{The quantum transport model adopted by this study is based on the mean field approximation. In addition, the proposed model is based on single-band effective mass approximation of the complex band structure of the material. A detailed discussion of the advantages and limitations of the adopted quantum transport model could be found in \cite{datta2012lessons}. However, the effective mass approximation is a computationally efficient method compared to other computationally intensive methods that account for the complex band structure of the material. The self-consistent solution of the magnetization dynamics, electric polarization, and quantum transport requires thousands of evaluations of the quantum transport model.}

The paper is organized as follows. Section \ref{ch_formalism} is dedicated to explaining the difference between MTJ, FTJ, and MFTJ along with the origin of TER and TMR effects. Section \ref{ch_dft} is devoted to DFT simulations of the ME effect at the FM/FE interface. The magnetization dynamics, ferroelectric dynamics, and quantum transport (NEGF) are illustrated in sections \ref{ch_LLG}, \ref{ch_LK}, and \ref{ch_NEGF}, respectively. \jpapT{Section \ref{ch_Perturbation} is dedicated to the formulation of the magnetic exchange coefficient as a function of the electric polarization based on time-dependent perturbation theory. The simulation procedure is explained in section \ref{ch_procedure}.} Finally, section \ref{ch_results} is assigned to the simulation results followed by conclusions in section \ref{conc}.

\section{Multiferroic Tunnel Junctions}\label{ch_formalism}

\sework{The} MFTJ structure combines the FM electrodes of MTJ together with the FE insulator of FTJ to \sework{produce} a four-state device. \rework{We} start by explaining the TMR effect in \sework{the} MTJ along with \sework{the} TER effect in FTJ before describing the MFTJ characteristics.  Furthermore, the ME effect at the FM/FE interface is a unique property of MFTJs that enhances the TER effect, as explained in detail in this section and \sework{in} section \ref{ch_dft}.

\begin{figure}[!ht]
\centering
\includegraphics[width=1\linewidth,trim={0cm 0cm 0cm 0cm}]{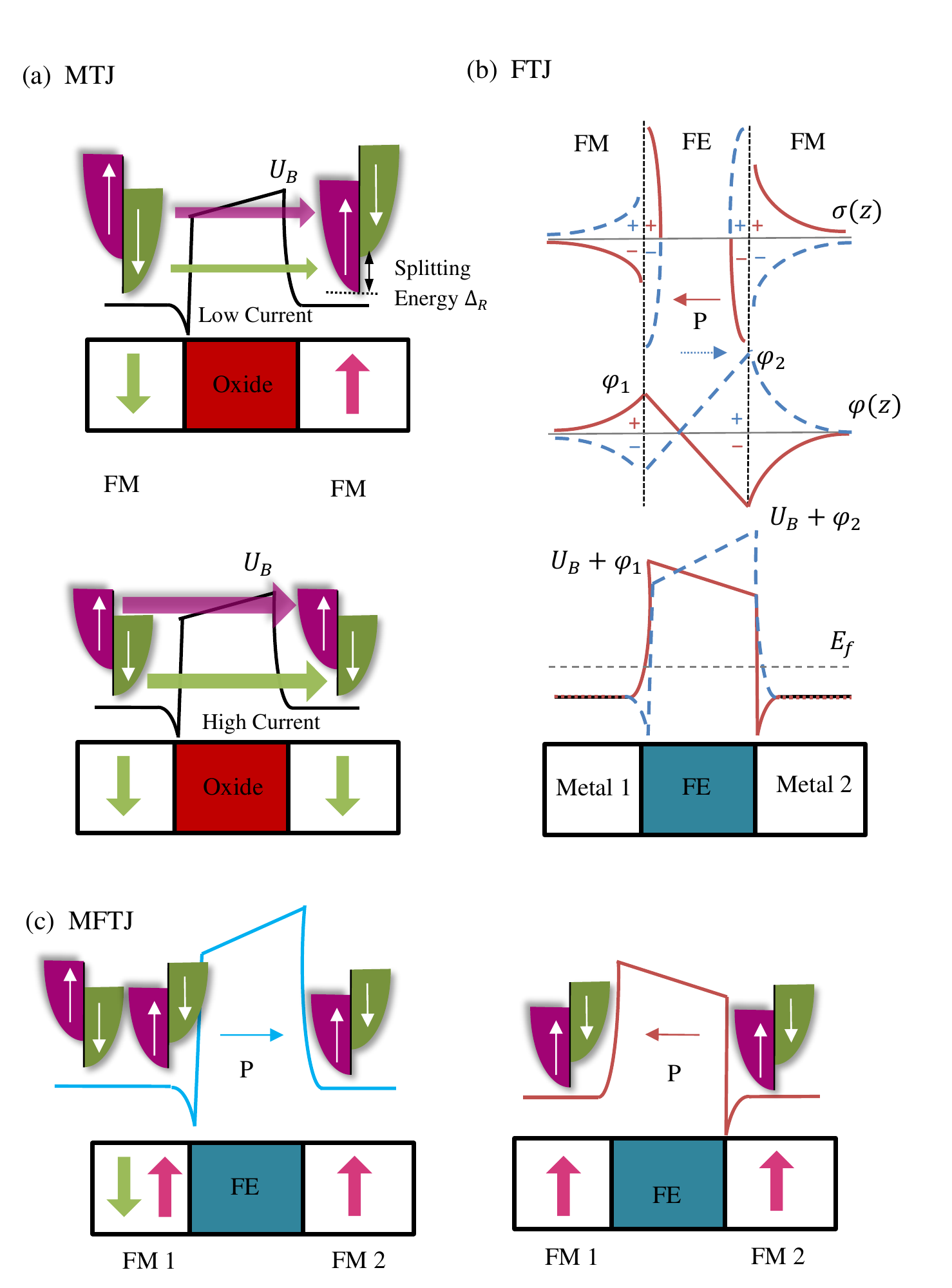}
\caption{The electrostatics of different tunnel junctions. (a) The electrostatic potential and density of states of MTJ. (b) The charge and electrostatic potential of FTJ. (c) The electrostatic potential and magnetization dependent density of states of MFTJ.}
\label{fig_MFTJ_band}
\end{figure}

The TMR effect could be explained in the light of spin dependent transport illustrated in Fig. \ref{fig_MFTJ_band} \cite{Julliere1975TunnelingFilms, Mathon2001TheoryJunction, Butler2001Spin-dependentSandwiches}. In such FM materials, the lower band of the density of states (DOS) of the majority and minority spin carriers have energy shift, as illustrated in Fig. \ref{fig_MFTJ_band} (a). The energy splitting is dependent on the magnetization direction. Therefore, in the case of anti-parallel alignment of the electrode magnetization, the majority spin carriers that migrate from the left electrode \rework{are restricted by the shortage} of matched spin states at the right electrode. Consequently, the overall charge current is reduced in the case of anti-parallel alignment of \sework{the electrode} magnetization. In the case of parallel alignment of the magnetization, the majority and minority spin carriers migrate from the left electrode \rework{and} are absorbed by the matched spin states that are sufficiently available at the right electrode. Consequently, the overall charge current is not limited by the availability of the spin states in the case of parallel alignment of magnetization. In other words, the MTJ resistance changes according to the magnetization alignment of the electrodes that control the DOS energy splitting between the majority and the minority spin carriers. Finally, the TMR is defined as $TMR=\frac{R_{AP}-R_{P}}{R_{P}}$, where $R_{AP}$ is the resistance of anti-parallel aligned magnetization state, and $R_P$ is the resistance of parallel aligned magnetization state \cite{Julliere1975TunnelingFilms}.

The TER effect of FTJ could be explained by the help of charge screening phenomena in the metal electrode \cite{Zhuravlev2005}. The electric polarization of FE insulator induces bound charges at the metal/FE interface. The bound charges are partially screened by the free electron gas in the metal side, as illustrated in Fig. \ref{fig_MFTJ_band} (b). The uncompensated charges at the interface result in a constant electric field inside the insulator, and hence linear potential. Moreover, the polarization direction controls the polarity of the bound charge, and hence the polarity of the potential drop in the FE insulator $(\phi_1-\phi_2)$, where $\phi_1$ and $\phi_2$ are the potential at the left and right metal/FE interfaces, respectively. Therefore, the barrier height increases by $|\phi_1+\phi_2|$ in the case of positive electric polarization.  In contrast, the barrier height is reduced by $|\phi_1+\phi_2|$ in the case of negative electric polarization. Finally, the large TER value of FTJ is a natural result of the exponential dependence of tunneling current on the barrier height. \jpapT{The asymmetry of the electrodes screening lengths is an important factor for MFTJs to exhibit a nonzero TER.  It is important to mention that the TER also depends on the barrier effective thickness, which can change upon the polarization reversal due to the change from the insulating to metal phase, and both interface terminations \cite{quindeau2015origin,borisov2015spin}.}

The aforementioned qualitative discussion could be formulated quantitatively, as illustrated in \cite{Zhuravlev2005}. The screening charges and potential distribution in the metal side is described by the Thomas-Fermi formalism:
\begin{eqnarray}
&\phi(z)= \left \{ \begin{array}{cc} 
\frac{\sigma_{s} \delta_1}{\epsilon_0\epsilon_{r1}}e^{-\frac{|z|}{\delta_1}}  & \quad  \quad z\leqslant 0 \\
\\
-\frac{\sigma_{s} \delta_2}{\epsilon_0\epsilon_{r2}}e^{-\frac{|z-d|}{\delta_2}}  & \quad  \quad z>0\sework{,} \\ \end{array} \right.\label{eqn_thomas_fermi}
\end{eqnarray}
\sework{where} $\phi$ is the electrostatic potential, $\sigma_s$ is the surface charge density of free charges, $\epsilon_0$ is permittivity of free space, \ze{$\epsilon_{r1}$ ($\epsilon_{r2}$) is the permittivity of the first (second) electrode, and $\delta_1$ ($\delta_2$) is the screening length of first (second) electrode}. \jpapT{According to Thomas-Fermi relation, the charge and the potential of any point in the electrodes decrease as an exponential function of the distance between the point and the interface.}  Moreover, the potential values at the interface are defined as $\phi_1=\frac{\sigma_{s}\delta_1}{\epsilon_0\epsilon_{r2}}$ and $\phi_2=\frac{-\sigma_{s}\delta_2}{\epsilon_0\epsilon_{r2}}$. By applying the Gauss's law at the metal/FE interface, we get the expression 
\begin{eqnarray}
\label{eqn_gauss}
& E_{FE} =\frac{(\sigma_s-P)}{\epsilon_0},
\end{eqnarray}
where $P$ is the polarization vector and $E_{FE}$ is the electric field in the FE layer. The potential drop $\phi_1-\phi_2$ is equal to the constant electric field inside the FE insulator multiplied by $t_{FE}$ as given by
\begin{eqnarray}
\label{eqn_pot}
&\frac{\sigma_{s}\delta_1}{\epsilon_0\epsilon_{r1}}+\frac{\sigma_{s}\delta_1}{\epsilon_0\epsilon_{r2}} + E_{FE}t_{FE}=0.
\end{eqnarray}
Finally, from (\ref{eqn_gauss}) and (\ref{eqn_pot}), the $\sigma_s$ that satisfies the continuity of the potential at the interface is defined as
\begin{eqnarray}
\label{eqn_sigma_s}
&\sigma_{s}=\frac{Pt_{FE} }{\frac{\delta_1}{\epsilon_{r1}}+\frac{\delta_2}{\epsilon_{r2}}+t_{FE}}.
\end{eqnarray}
However, for the limiting case of $t_{FE}\gg \frac{\delta_1}{\epsilon_{r1}}+\frac{\delta_2}{\epsilon_{r2} }$, the free charge density $\sigma_s$ is equal to $P$, and hence the potential drop is zero which \rework{eliminates} the TER effect \ze{\cite{Zhuravlev2005}}. Therefore, a mandatory constraint is imposed on the maximum FE thickness that maintains the TER effect. However, the stability of the ferroelectricity imposes a lower limit on the FE thickness. Therefore, the FE layer \sework{should have} an optimal thickness that maintains the ferroelectricity and provides high TER ratio at the same time.

An MFTJ combines \sework{the} TER effect of FTJ along with \sework{the} TMR effect of MTJ to produce a four-state device as illustrated in Fig. \ref{fig_MFTJ_band} (c). However, it has been experimentally observed \rework{that} the LCMO electrode of LSMO/LCMO/BTO/LSMO MFTJ \cite{Yin2015} goes through phase transition from the FM state to the AFM phase \sework{by the influence of the electric polarization switching. To understand the effect of the FM to the AFM phase transition on the TER, let us assume that both electrodes have the magnetization in the positive $z$ direction. In the case of \textit{positive} electric polarization (\textit{high resistance state}), the LCMO left electrode has an \textit{AFM} configuration. The lower band of the DOS of the spin-up carriers shifts to higher energy levels. Therefore, the spin-up carriers that migrate from the right electrode are restricted by the shortage of spin-up states at the left electrode. Thus, the MFTJ \textit{high resistance} increases, and hence the TER increases. In case of \textit{negative} electric polarization (\textit{low resistance state}), the LCMO has an \textit{FM} configuration. The spin-up carriers that migrate from the right electrode are absorbed by the matched spin states without any restriction. Consequently, the overall TER of the MFTJ improves.} The details and origin of this ME effect are explained in the following section.

\sework{\section{The First Principle Calculations of the FM/FE interface}}\label{ch_dft} 
\sework{Recently, many different forms of magnetoelectric effects have been observed in the literature \sework{such as} electric field manipulated magnetization, electric field induced magnetic phase transition, and voltage controlled magnetic anisotropy \cite{weisheit2007electric,maruyama2009large, Eerenstein2006MultiferroicMaterialsb}}.
In this study, we focus on the magnetoelectric effect that happens in the interface between $La_{1-x}A_{x}MnO_3/BaTiO_3$, where A is a divalent cation, i.e., Ca, Ba, and Sr and $x$ is the chemical doping concentration.
$LAMO$'s phase diagram exhibits a phase change between the ferromagnetic state and antiferromagnetic state as a function of hole carrier concentration $x$ \cite{dagotto2001colossal}. The transition between the FM and AFM phases and its dependence on hole concentration could be explained by the existence of two competing interactions that happen between the adjacent Mn sites in LAMO: superexchange interaction and double exchange interaction. In contrast to superexchange interaction that prefers AFM alignments, the double exchange interaction favors FM alignment \cite{de1960effects}. 
The doping concentration $x,$ modulates the density of electrons in Mn $e_g$ orbitals that mediate the double exchange interaction. Note, the doping concentration supports one of the interactions over the other, and hence favors one of the configurations over the other. 

\begin{figure}[!h]
\centering
\includegraphics[width=1\linewidth,trim={.3cm 0cm 1cm 0cm},clip=true]{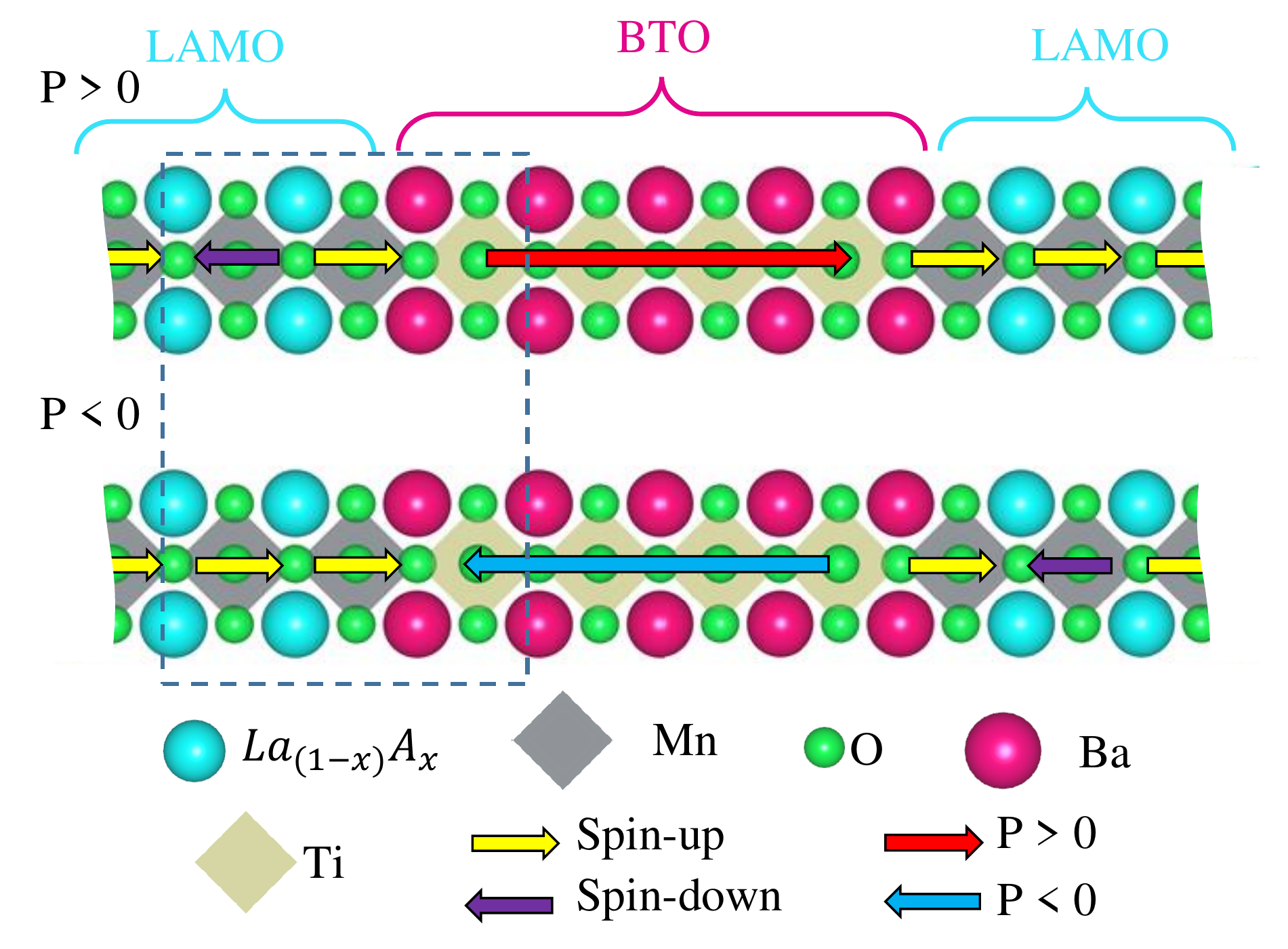}
\caption{The atomic structure of LAMO/BTO interface that is used for supercell simulation. The structure consists of 4.5 unit cells of LAMO and 5.5 unit cells of BTO. \me{The dependence of the magnetic configuration of LAMO on the electric polarization of BTO is illustrated.} At the left interface, the second Mn site in LAMO exhibits AFM and \textit{FM} alignment in case of positive and \textit{negative} electric polarization of BTO, respectively.}
\label{fig_lsmo_bto}
\end{figure}

The electrostatic doping created by electric polarization could change the electron concentrations similar to chemical doping \cite{ahn2003electric}. Since the bound charges induced by electric polarization of BTO at the interface modulates the screening charges at the LAMO side, the electric polarization could control the magnetization phase transition similar to chemical doping. \ze{The magnetoelectric effect in LAMO/BTO interface is illustrated in Fig. \ref{fig_lsmo_bto}. The second Mn site in LAMO exhibits AFM (\textit{FM}) alignment in case of positive (\textit{negative}) polarization state.} However, the chemical doping concentration has to be fixed at \rework{the magnetic phase transition point ($x=0.5$) between the FM and the AFM phases} to facilitate the magnetic phase transition by electrostatic doping \cite{akimoto1998antiferromagnetic}.

\subsection{Simulation Procedure and Parameters}

We applied DFT method to extract the electrostatic potential profile of LAMO/BTO and Co/BTO structures. The screening lengths of LAMO and Co electrodes are estimated from the electrostatic potential. The generalized gradient approximation (GGA) method \cite{perdew1996generalized} implemented in Quantum-ESPRESSO package \cite{giannozzi2009quantum} is used to perform all of the DFT calculations in this study.  The Vanderbilt’s ultrasoft pseudopotential \cite{vanderbilt1990soft} is used along with virtual crystal approximation (VCA) \cite{Colombo1991Valence-bandInterfaces} to handle the La-A doping. The VCA method is used by Burton \textit{et al.} \cite{burton2009prediction} to perform DFT calculations for typical structure with acceptable accuracy. The energy cutoff of 400 eV and Monkhorst-Pack grid of 12x12x1 of k-points are used for all the DFT simulations in this study.

\begin{figure}[!h]
\centering
\includegraphics[width=1\linewidth,trim={0cm 0cm 0cm 0cm},clip=true]{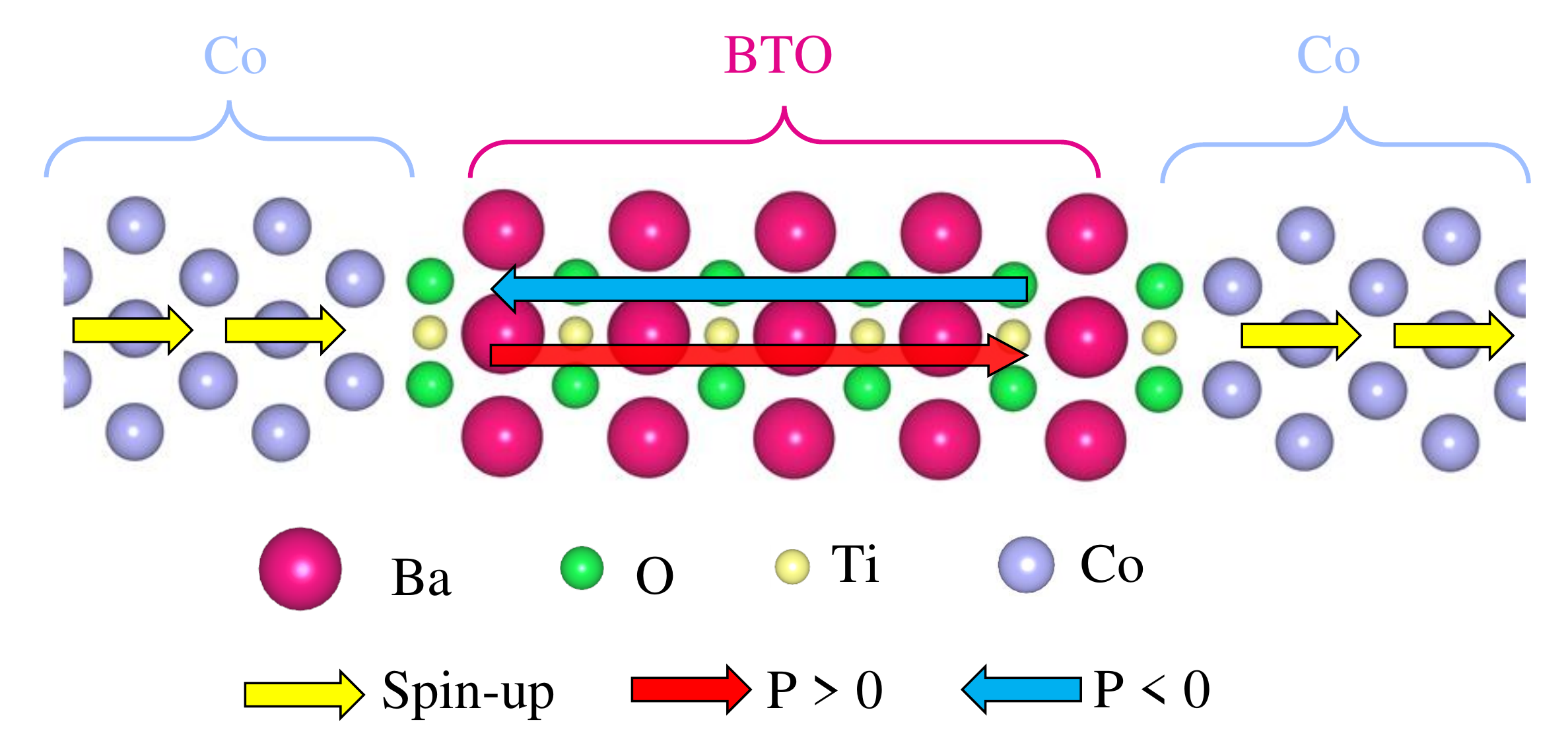}
\vspace{-2\baselineskip}
\caption{The atomic structure of Co/BTO interface that is used for supercell simulation. The structure consists of $5.5$ unit cells of BTO and $4.5$ unit cells of Co. The Co magnetization exhibits constant magnetization independent of the electric polarization of BTO.}
\label{fig_co_bto}
\end{figure}

\jpapT{The supercell used to simulate LAMO/BTO interface consists of 4.5 unit cells of LAMO and 5.5 unit cells of BTO, as illustrated in Fig. \ref{fig_lsmo_bto}. The structure is stacked along (001) direction of the perovskite cell. The stacking sequence at LAMO/BTO interface is $AO-BO_2$ \cite{burton2009prediction}. The supercell illustrated in Fig. \ref{fig_co_bto} is utilized to model the Co/BTO interface. The structure consists of 4.5 unit cells of Co and 5.5 unit cells BTO along (001) direction and is rotated $45^\circ$ in the $x-y$ plane. The most stable interface has TiO$_2$ termination as described in \cite{Cao2011}. We did not include any vacuum regions in these structures. As both structures are epitaxial growth on a SrTiO$_3$ substrate that has a bulk in-plane lattice constant of $a = 3.94\AA$, the lattice constant in the lateral direction is constrained to $a= 3.94\AA$ for all the layers of LAMO, Co and BTO. The lateral strain results in tetragonal distortion in the longitudinal direction ($z$ direction). To estimate the tetragonal distortion, the DFT calculation of a single LAMO unit cell is repeated with different longitudinal lattice constants $c$. The lattice constant that has the minimum total energy is selected for further supercell simulations. The longitudinal lattice constant of LAMO that has minimum total energy is $c/a = 0.99$. Similarly, the longitudinal lattice constant of BTO and Co are estimated to be $c/a = 1.05$ and $c/a = 0.83$, respectively. Next, both supercells of LAMO/BTO and Co/BTO with the in-plane constraint and the corresponding tetragonal distortion are relaxed until the total force on the atoms is less than $10^{-3}$ $Ryd/au$. As the total force on the atoms reaches the limit of $10^{-3}$ $Ryd/au$, the atoms reach their equilibrium positions. Further optimization beyond this limit results in a negligible change in the positions of the atoms.}

\begin{figure}[!ht]
\centering
\includegraphics[width=1\linewidth,trim={0cm 0cm 0cm 0cm},clip=true]{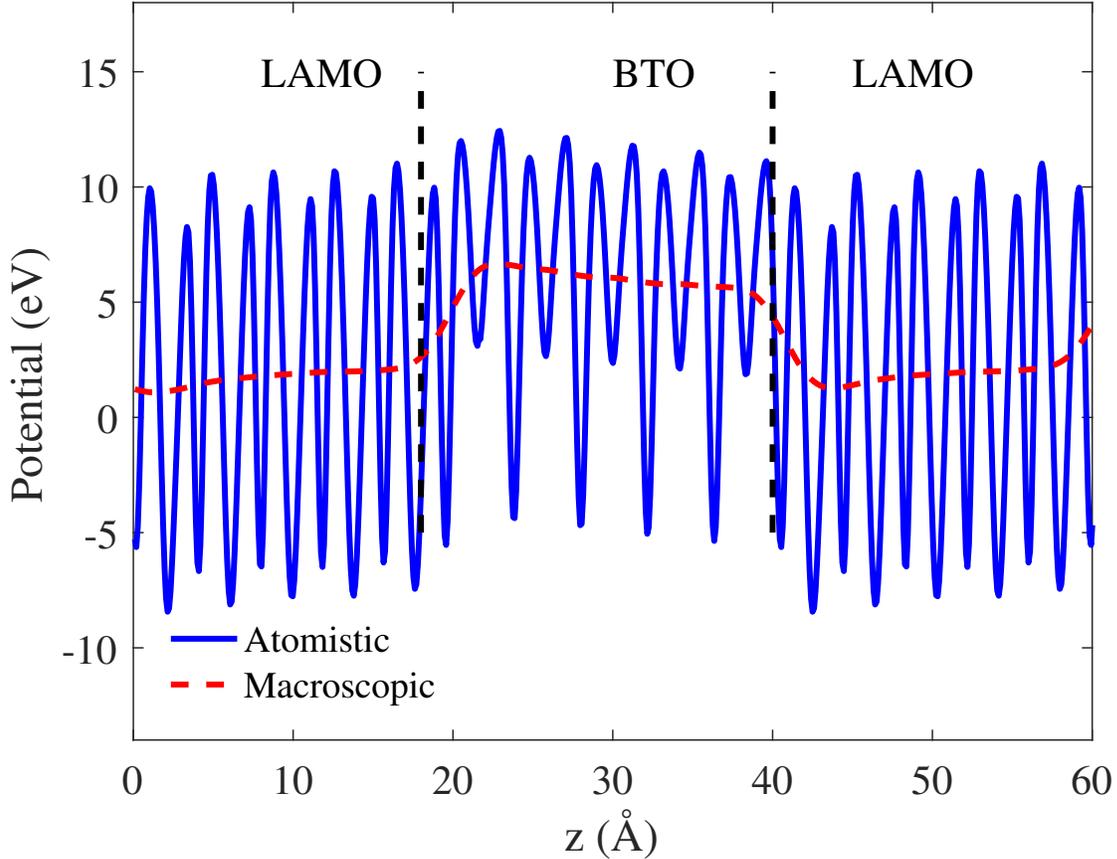}

\caption{The atomistic and macroscopic potential of LAMO/BTO structure.}
\label{fig_lsmo_bto_pot}
\end{figure}

\begin{figure}[!ht]
\centering
\includegraphics[width=0.9\linewidth,trim={0cm 0cm 0cm 0cm},clip=true]{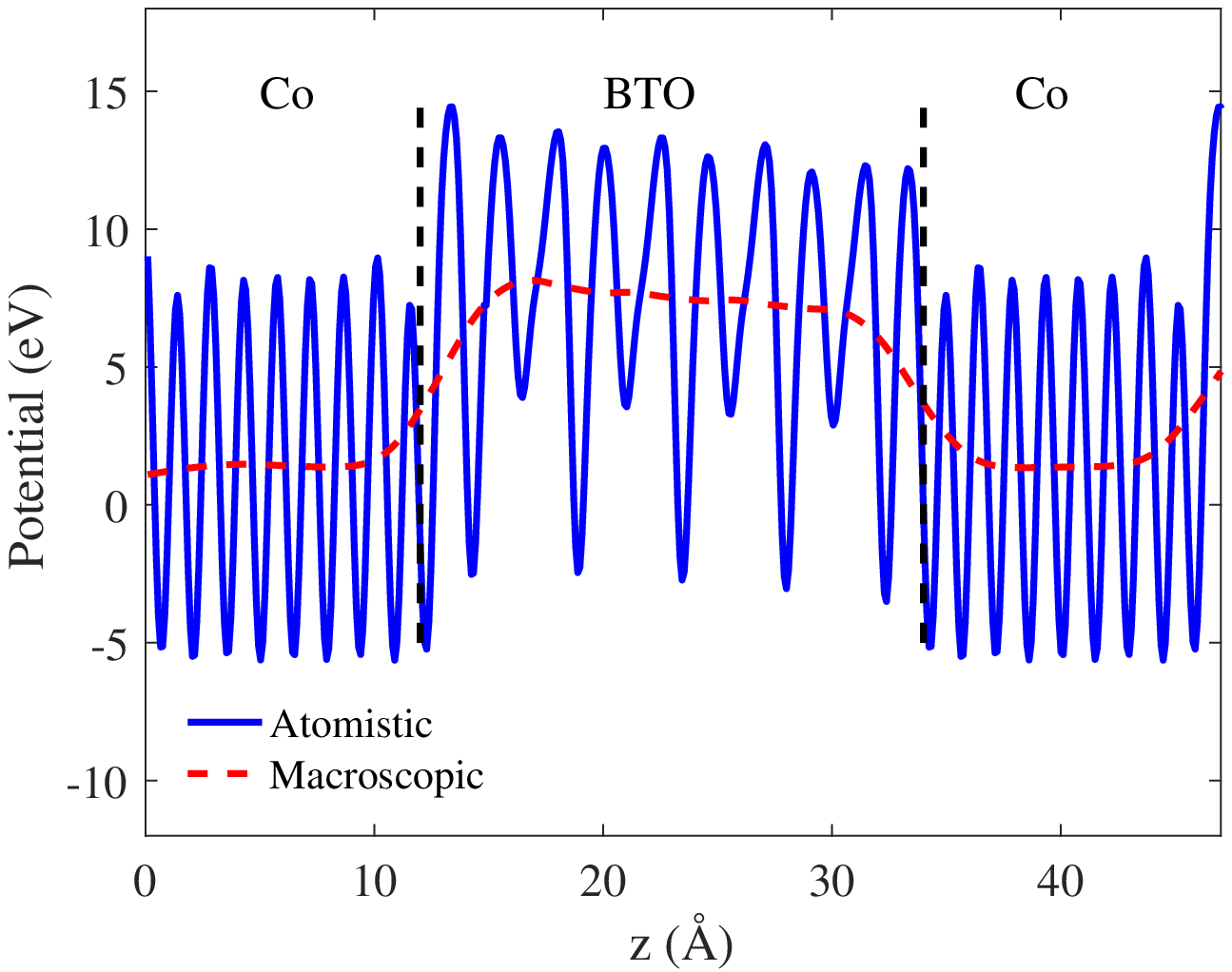}

\caption{The atomistic and macroscopic potential of Co/BTO structure.}
\label{fig_co_bto_pot}
\end{figure}

\subsection{DFT Simulation Results}
\jpapT{The magnetic configurations that minimize the total energy of the LAMO/BTO interface are illustrated in Fig. \ref{fig_lsmo_bto}. The second Mn site in LAMO (left interface) exhibits AFM (\textit{FM}) alignment in case of positive (\textit{negative}) polarization state. The magnetic configuration of LAMO (left interface), that minimize the total energy, is AFM in the case of positive polarization and FM in the case of negative polarization \cite{burton2009prediction}. The DFT simulation results for the magnetization of Mn sites are 2.57, -2.66, 2.76, 2.87, and 3.02 $\mu_B$. In case of positive polarization state, the magnitude of the magnetization of the Mn atoms is lower at left interface and increases for the atoms away from that left interface \cite{burton2009prediction}.  In contrast, the magnetic configuration of the Co/BTO interface, that minimizes the total energy, is FM configuration independent of electric polarization of BTO. The Co/BTO interface exhibits a constant magnetic configuration independent of the electric polarization of BTO \cite{Cao2011}, as illustrated in Fig. \ref{fig_co_bto}. The DFT simulation result for the magnetization of bulk Co is $1.74$ $\mu_B$.}


\jpapT{The atomistic electrostatic potential of LAMO/BTO and Co/BTO are illustrated in Fig. \ref{fig_lsmo_bto_pot} and Fig. \ref{fig_co_bto_pot}, respectively. The macroscopic potential is estimated from the atomistic potential by a moving window integral \cite{Colombo1991Valence-bandInterfaces} and fitted by a spline function. Finally, the screening lengths $\delta/\epsilon_0\epsilon_r$ of La$_{0.7}$A$_{0.3}$MO, L$_{0.5}$A$_{0.5}$MO, and Co are estimated from electrostatic potential to be 1.06, 1.05, and 1.14 $m^2/F$, respectively. The estimated screening lengths are used in the spin dependent transport calculations, as explained in section \ref{ch_NEGF}.}

\section{The Magnetization Dynamics}\label{ch_LLG}

The Landau-Lifshitz-Gilbert (LLG) equation formulates the precessional and damped motion of magnetization induced by the magnetic field and spin current \cite{Landau1935OnBodies, Landau1935OnBodies,dAquino2005NonlinearNanoparticles}. The single domain LLG equation is used along with NEGF self-consistently in many studies in literature. The single domain solution of LLG equation is not appropriate for LAMO Because the LAMO material has atoms in the FM order and other atoms in the AFM order at the same time. Similarly, the solution of the magnetization as a continuum fails as well, because it requires a second order derivative of the magnetization with respect to space. The second order derivative appears in the definition of the exchange interaction effective field $H_{exch}=\frac{2A}{\mu_0 M_S}\frac{\partial^2 m}{\partial x^2}$, where $A$ is a constant. However, the abrupt change of the magnetization at that atomistic scale makes the derivative with respect to space is not possible. The usual solution in case of AFM material is to replace the magnetization by the total magnetization $l=m_{j+1}+m_j$ and the AFM Neel field $n=m_{j+1}-m_j$ that are continuous variables in case of AFM material. However, LAMO has the FM and AFM orders that exist at the same time. The Neel field will be discontinuous at the area between the FM and the AFM phase. 

\jpapT{We adopted a discrete multi-domain version of the LLG equation. The main difference between the continuum and discrete LLG equation is the definition of the exchange field. The definition of the exchange field in the discrete LLG equation does not require differentiation with respect to spatial coordinates. The discrete multi-domain LLG equation is similar to the atomistic LLG equation \cite{evans2014atomistic}. However, the discrete multi-domain LLG equation models the lateral direction as a single domain to reduce the computational effort. The lateral single domain assumption does not affect the accuracy of the method because the cross-section area of the junction is large enough to neglect the effect of edges. The discrete domains have a thickness equal to a single unit cell in the normal direction.}

\subsection{The LLG Equation}

The LLG equation \cite{evans2014atomistic} can be expressed as
\begin{eqnarray}
\label{eqn_LLG}
\frac{\partial m_i}{\partial \tau} &=& - m_i \times H_{eff,i} - \alpha m_i \times m_i \times H_{eff,i} +\nonumber\\
&& STT_i,
\end{eqnarray}
where $m$ is a unit vector in direction of magnetization, $\tau$ is defined as $\tau=\frac{|\gamma|}{(1+{\alpha}^{2})}dt$, t is the time, $H_{eff}$ is the effective magnetic field, $\alpha$ is the Gilbert damping constant, $\gamma$ is the gyromagnetic ratio,i is index over the atoms along the $x$ axis, and STT is the spin transfer torque. The first term of (\ref{eqn_LLG}) is the precessional motion of the magnetization due to the effective magnetic field. The second term models the damped part of magnetization oscillation. The third term is the spin torque exerted by the spin current on the magnetization.

\jpapT{LAMO is modeled as a 1D chain of discrete domains with a magnetization variable $m_i$ assigned to each domain.} Each mesh cell has a length equal to the lattice constant and cross section area equals the total cross-section area of the MFTJ.  In other words, we assumed that the MFTJ cross-section area is large enough. Therefore, we can neglect the effect of the boundary cell on the magnetization dynamics.

\subsection{The Effective Magnetic Field}

The effective magnetic field is given by
\begin{eqnarray}
H_{eff,i}=H_{ext}+H_{therm}+H_{anis} +H_{exch,i},\label{eqn_H_eff}\label{eqn_H_anis}
\end{eqnarray}
where $H_{ext}$ is the external magnetic field, $H_{exch,i}$ is the exchange interaction effective field, $H_{therm}$ models the random thermal variations, and $H_{anis}$ is the magnetic anisotropy. The magnetic anisotropy is defined as $H_{anis}=\frac{2K_{U}}{M_s}$, where $K_{U}$ is the anisotropy constant, and $M_S$ is the saturation magnetization.

The exchange interaction field is defined as $H_{exch,i}=\frac{1}{\mu_0 M_S}\sum_j^N J'_{exch,i,j} m_j$ \cite{evans2014atomistic}, where $J'_{exch,i,j}$ is the material magnetic exchange coefficient that is averaged by the FM to AFM transition probability as explained in \ref{ch_Perturbation}, and N is the number of nearest neighbors.

\subsection{The Thermal Fluctuations}

The random variation in the magnetization due to thermal excitations is modeled by the effective magnetic field $H_{therm}$ defined as $H_{therm}= \zeta \sqrt[]{\frac{2\alpha KT }{|\gamma| M_{S} V_{cell} dt }}$, where $K$ is Boltzmann’s constant, $T$ is the temperature, $V_{cell}$ is the mesh cell volume that equals to the lattice constant multiplied by the total cross-section area of the MFTJ, $dt$ is the numerical time step, and $\zeta$ is a vector with random components which are selected from standard normal distribution \cite{BrownJr1963ThermalParticleb}.

The thermal fluctuations term makes the LLG equation stochastic differential equation (SDE). The integration of the thermal fluctuations results in Wiener stochastic process \cite{berkov2007magnetization, gardiner1986handbook} that is not differentiable with respect to time. Therefore, the Stieltjes integral is used instead of the Riemann integral to integrate the thermal term \cite{berkov2007magnetization, gardiner1986handbook}. The Stieltjes integral of the thermal term is defined in terms of the differential increments of Wiener process that has a variance proportional to the integration time step. Therefore, time-step $dt$ appears in the denominator of the thermal field. The details of the integration of the LLG equation as a stochastic differential equation is explained in \cite{scholz2000langevin,berkov2007magnetization,evans2014atomistic}.

\subsection{The spin transfer torque}
 The STT term is defined as $STT=\frac{\hbar }{2 \mu_0 M_s a} m\times ( m\times J^{spin})$ \cite{Datta2012Voltage}, where $a$ is lattice constant, and $J^{spin}$ is the spin current that is calculated from the quantum transport, as explained in \ref{ch_NEGF}.
The definition of STT term adopted in this paper is preferred over the Slonczewski STT term in the context of the quantum transport. The Slonczewski STT term has parameters that depend on the material and geometry of the junction to account for the junction efficiency of producing spin current. These effects are already included in the quantum transport. Therefore, we avoided using the Slonczewski STT term in favour of the quantum transport formulation of the spin current.

\section{The Ferroelectric Dynamics} \label{ch_LK}

The Landau-Devonshire (LD) expression of the free energy of FE material \rework{that} describes the dependence of free energy of the FE material on the electric polarization and the electric field is \sework{defined as}
\begin{eqnarray}
F=&\alpha_1 P^2+\alpha_{11}P^4+\alpha_{111}P^6-&\frac{V_aP}{t_{FE}}\label{eqn_LD},
\end{eqnarray}
\sework{where} $\alpha_{1}$, $\alpha_{11}$, and $\alpha_{111}$ are the free-energy expansion coefficients for bulk material \cite{liu2013effect,tan2001theory}. The polarization of the material could be determined by minimizing the free energy (F) with respect to the electric polarization in (\ref{eqn_LD}).
However, the LD expansion describes the static relation between electric field and polarization. The dynamic behavior of FE and its dependence on time is described by Landau-Khalatnikov (LK) \sework{equation:}
\begin{eqnarray}
\lambda \frac{\partial P}{\partial t}=-\frac{\partial F}{\partial P},\label{eqn_LK}
\end{eqnarray}
\sework{where} $\lambda$ is the viscosity coefficient that represents the resistance of FE polarization motion toward the free energy minimum state.

\sework{\section{Quantum Transport: Non-equilibrium Green's Function modeling of MFTJ} \label{ch_NEGF}}

\begin{figure}[!h]
\centering
\includegraphics[width=1\linewidth,trim={0cm 1cm 0cm 0.5cm},clip]{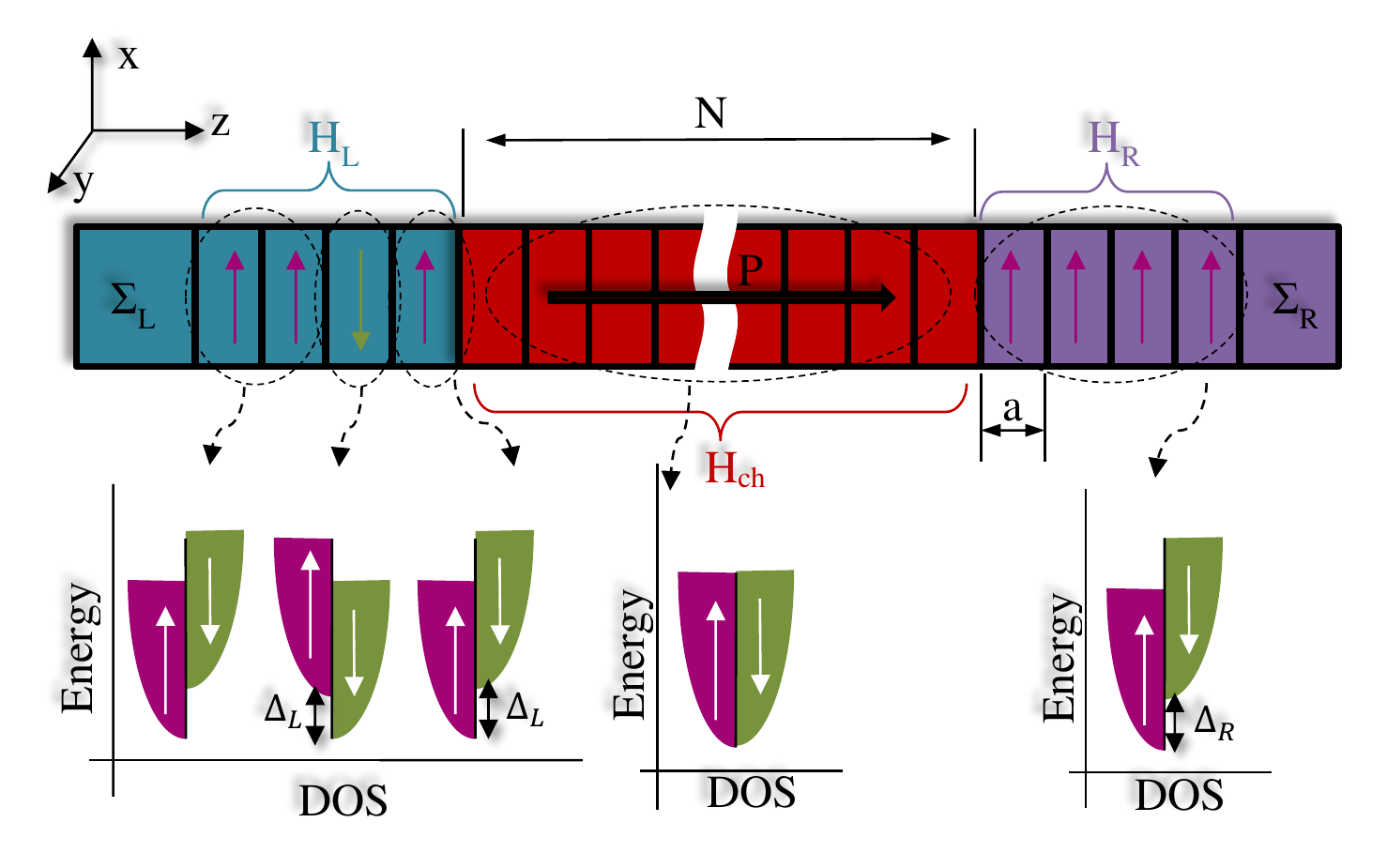}
\caption{The MFTJ structure with the spin based NEGF meshing projected. The Hamiltonian definitions along with magnetization directions and DOS.}
\label{fig_negf}
\end{figure}

The NEGF models the magnetization dependent tunneling current by splitting the device into two independent channels for the spin-up and spin-down carriers. The schematic diagram in Fig. \ref{fig_negf} shows the device meshing and the magnetization dependent DOS. The spin-based channel Hamiltonian $H_{ch}$ and the left (right) contact Hamiltonian $H_{L(R)}$ \cite{Datta2012Voltage,Datta2010QuantitativeDevices} of the MFTJ are defined as 
\begin{widetext}
\begin{eqnarray}
\label{eqn_h_ch}
H_{L(R)}&=& \left \{ \begin{array}{cc} 
\left(\alpha_{L(R)} \pm \frac{qV_a}{2}\right)I \pm \left(I-\sigma. M_{L(R)}(i)\right)\frac{\Delta_{L(R)}}{2}, \qquad \quad \quad & i=j  \\
\\
-t_{L(R)} I, \quad\quad\quad &  j=i \pm 1 \\ 
\\
0,\quad \quad \quad & o.w. \end{array} \right.\\
H_{ch}&=& \left \{ \begin{array}{cc} 
(\alpha_{ch}+U_B)I+(qV_a+\phi_{BI}+\phi_1-\phi_2)(\frac{N+1-2i}{2N+2})I, \quad \quad  &i=j  \\
\\
-t_{ch} I,   \quad \quad &j=i \pm 1 \\ 
\\
0, \quad \quad & o.w. \end{array} \right.\label{eqn_h_LR}
\end{eqnarray}
\end{widetext}
where $a$ is the length of the mesh element, N is the number of the mesh elements, q is the electron charge, $V_a$ is the applied voltage, $\phi_{BI}$ is the built-in potential, I is the identity matrix, i is the horizontal index of the Hamiltonian matrix, j is the vertical index of the Hamiltonian matrix, $\sigma$ are the Pauli spin matrices, $\Delta_{L(R)}$ is the splitting energy of the left (right) contact as illustrated in Fig. \ref{fig_negf}, respectively, and $M_{L(R)}$ \sework{is the normalized magnetization of the left (right) contact.} 
The Hamiltonian tight binding parameters are defined as $\alpha_{ch}(K_t)=2t_{ch}+\frac{\hbar^2K_t^2}{2m_{ch}^*}$, $\alpha_{L(R)}(K_t)=2t_{L(R)}+\frac{\hbar^2K_t^2}{2m_{L(R)}^*}$, $t_{ch}=\hbar/(2m_{ch}^*a^2)$, and $t_{L(R)}=\hbar/(2m_{L(R)}^* a^2)$, where $K_t$ is the momentum vector in the transverse direction, $m_{L(R)}^*$ \sework{is the electron effective mass of left (right)} electrodes, and $m_{ch}^*$ is the electron effective mass of the channel. 

The Hamiltonian $H_{L(R)}$ dependence on the magnetization direction is modeled by the term $\left(I-\sigma. M_{L(R)}(i)\right)\frac{\Delta_{L(R)}}{2}$. The term $(qV_a+\phi_{BI}+\phi_1-\phi_2)(\frac{N+1-2i}{2N+2})I$ in $H_{ch}$ linearly interpolates the applied field and built-in potential $V_a+\phi_{BI}+\phi_1-\phi_2$  over the channel. $U_B$ is the barrier height relative to the conduction band. The term $\phi_1-\phi_2$\ze{,} that is estimated by (\ref{eqn_thomas_fermi}-\ref{eqn_sigma_s}), represents the dependence of the potential on the electric polarization of the FE insulator. The screening lengths estimated by the DFT method and the electric polarization \ze{estimated} by LK equation are plugged in Thomas-Fermi relation (\ref{eqn_thomas_fermi}) to \ze{determine} $|\phi_1-\phi_2|$. \ze{The term $|\phi_1-\phi_2|$ is necessary for evaluating the NEGF Hamiltonian (\ref{eqn_h_ch}).}

\sework{The Green's function $G$ is defined as}
\begin{eqnarray}
&G=\left[EI-H-\Sigma_L-\Sigma_R\right]^{-1}, \label{eqn_g}
\end{eqnarray}
\sework{where $H$ is the full device Hamiltonian, $\Sigma_{L(R)}$ is the left (right) contact self-energy that is defined as}
\begin{eqnarray}
&\Sigma_{L}=
\begin{bmatrix}
-t_{L}e^{iK^\uparrow_{L} a}&0&0&.&.&0&\\
0&-t_{L}e^{iK^\downarrow_{L} a}&0&.&.&.&\\ 
0&0&.&.&.&.&\\
.&.&.&.&.&.&\\
.&.&.&.&.&.&\\
0&.&.&.&.&0&\\\
\end{bmatrix},\\
&\Sigma_{R}=
\begin{bmatrix}
0&.&.&.&.&0&\\
.&.&.&.&.&.&\\
.&.&.&.&.&.&\\
.&.&.&.&0&.&\\
.&.&.&0&-t_{R}e^{iK^\uparrow_{R} a}&0\\
0&.&.&.&0&-t_{R}e^{iK^\downarrow_{R} a}\\
\end{bmatrix},\label{eqn_self_energy}
\end{eqnarray}
\sework{where $K^\uparrow_{L(R)}$ is the left (right) contact longitudinal wave vector of \textit{spin-up} electron given by}
\begin{eqnarray}
&K^\uparrow_{L(R)}=cos^{-1} \left(1-\frac{E \pm \frac{qV_a}{2}-\frac{h^2K^2_t}{2m_{L(R)}^*}+\frac{\Delta_{L(R)}}{2}}{2t_{L(R)}} \right),
\end{eqnarray}
and $K^\downarrow_{L(R)}$ is the left (right) contact longitudinal wave vector of \textit{spin-down} electron given by
\\
\begin{eqnarray}
&K^\downarrow_{L(R)}=cos^{-1} \left(1-\frac{E \pm \frac{qV_a}{2}-\left(\frac{h^2K^2_t}{2m_{L(R)}^*}\right)-\frac{\Delta_{L(R)}}{2}}{2t_{L(R)}} \right).
\end{eqnarray}
\sework{Finally, the Landau's current formula is defined as}
\begin{eqnarray}\label{eqn_current}
J=\frac{-e}{2\pi^2h}\int^{\infty}_{-\infty}\int^{\infty}_{-\infty}dk_xdk_y\int dE t\left(f_L-f_R\right),
\end{eqnarray}
\sework{where t is the transmission coefficient of the channel given by the expression}
\begin{eqnarray}\label{eqn_trans}
t=Trace\left(\Gamma_L G \Gamma_R G^\dagger \right),
\end{eqnarray}
and  $\Gamma_{L(R)}$ is the left (right) broadening function defined by
\begin{eqnarray}
&\Gamma_{L(R)}=i\left(\Sigma_{L(R)}-\Sigma^{\dagger}_{L(R)}\right).  \label{eqn_broadening}
\end{eqnarray}
The Fermi-Dirac distribution $f_L(R)$ is defined as
\begin{eqnarray}
f_{L(R)}=\frac{1}{e^{(E-\mu_{L(R)})/{K_BT}}+1}.
\end{eqnarray}

The spin current is defined as \cite{Datta2010QuantitativeDevices}

\begin{eqnarray}
J^{spin}=&&\frac{i}{ 2\pi^2 h}\int{dk_xdk_y}\int Trace[\sigma.(HG^n-...\nonumber\\ 
&&G^nH)_{j,j}]dE,\label{eqn_j_spin}
\end{eqnarray}
where Gn defined as 
\begin{eqnarray}
G^n=G(\Gamma_L f_L+\Gamma_R f_R)G^{\dagger}.
\end{eqnarray}

\section{Time-dependent Formulation of Exchange interaction Coefficient based on Time-dependent Perturbation Theory}\label{ch_Perturbation}
In the following discussion, we formulate a time-dependent formulation of the evolution from the FM to AFM phase induced by electric polarization. The sign of exchange interaction constant $J_{exch,i,j}$ is responsible for the magnetic order. The FM to AFM phase transition is controlled by the electrical polarization of the BTO. The change in the potential results from polarization switching can be considered as a small perturbation. Therefore we can use the time-dependent perturbation theory to model the time evolution of the exchange coefficient due to the perturbation potential \cite{griffiths2018introduction}. The time-dependent perturbation potential can be formulated from the Thomas-Fermi relation as 

\begin{eqnarray}
V(z,t)&=& 
\frac{P(V_a,t)t_{FE} }{\frac{\delta_1}{\epsilon_{r1}}+\frac{\delta_2}{\epsilon_{r2}}+t_{FE}}\frac{\delta_1}{\epsilon_0\epsilon_{r1}}e^{-\frac{|z|}{\delta_1}}-V_0,\\
V_0&=&\frac{P(V_a,t=0)t_{FE} }{\frac{\delta_1}{\epsilon_{r1}}+\frac{\delta_2}{\epsilon_{r2}}+t_{FE}}\frac{\delta_1}{\epsilon_0\epsilon_{r1}}e^{-\frac{|z|}{\delta_1}},
\end{eqnarray}
where $V_0$ is the initial value of the potential. Assuming that the FM to AFM phase transition results from the transition from wave function $\psi_b$ to $\psi_a$. The time dependent wave function can be written as $\Psi(z,t)=a(t)\psi_a(z)e^{-i \frac{E_a}{\hbar} t}+b(t)\psi_b(z)e^{-i \frac{E_b}{\hbar} t}$. The time evolution of a(t) can be formulated as \cite{griffiths2018introduction}

\begin{eqnarray}
a(t)&=&-\frac{i}{\hbar}\frac{t_{FE} }{\frac{\delta_1}{\epsilon_{r1}}+\frac{\delta_2}{\epsilon_{r2}}+t_{FE}}\frac{\delta_1}{\epsilon_0\epsilon_{r1}}\left<\psi_a| e^{-\frac{|z|}{\delta_1}}| \psi_b \right>\times ...\nonumber\\ 
&&\int^t_0(P(t)-P(t=0))e^{- i \omega t}dt,\label{eqn_a_t}
\end{eqnarray}
where $\omega=\frac{E_b-E_a}{\hbar}$, $E_a$ is the energy of $\psi_a$, and $E_b$ is the energy of $\psi_b$. finally, the transition probability from state $\psi_a$ to state $\psi_b$ is $P_{b\rightarrow a}=|a(t)|^2$. The term $\left<\psi_a|e^{-\frac{|z|}{\delta_1}}| \psi_b \right>$ can be determined from the normalization of the probability density. However, the polarization as function of time P(t) is not known analytically. Therefore the integration (\ref{eqn_a_t}) has to be numerically evaluated. The time evolution of the exchange interaction coefficient can be formulated as 
\begin{eqnarray}
J'_{exch,i,j}=|J_{exch,i,j}|\left(1-|a(t)|^2\right)-...\nonumber\\
|J_{exch,i,j}||a(t)|^2, \label{eqn_Jprim}
\end{eqnarray}
where $J_{exch,i,j}$ is the material-dependent magnetic exchange constant. 
\section{Simulation Procedure and Parameters selection}\label{ch_procedure}
In the previous discussion, we explained the quantum transport model, the magnetization LLG equation, and the LK equation, separately. In the following discussion, we explain the methodology we used to solve these models together to get the MFTJ characteristics. The steady-state characteristic of the MFTJ is calculated by the following procedure. Given the initial polarization $P$ and the external applied voltage $V_a$, the term $\frac{dF}{dP}$ is calculated by differentiating the LD equation analytically. The $\frac{dF}{dP}$ obtained in the previous step is substituted in the LK equation to get $\frac{dP}{dt}$. Then the forward difference formula is used to update the polarization $P(t+dt)=P(t)+\frac{dP}{dt}dt$. The previous steps are repeated iteratively until the electric polarization reaches its steady-state value. After the steady-state electric polarization is obtained, the electrostatic potential is calculated from (1) and (4). Next the current is calculated from the quantum transport model. However, the solution of the transport model is dependent on the magnetization directions of the electrodes which are calculated by the LLG equation. At the same time, the solution of the LLG equation depends on the current obtained from the NEGF equation. This raises the need for a self-consistent solution of the quantum transport and the LLG equation iteratively until the steady state current and magnetization are reached. 
The LLG equation is solved using Huen's method \cite{BrownJr1963ThermalParticleb, evans2014atomistic}. The existence of the thermal fluctuation in the LLG equation makes it a stochastic differential equation (SDE). The integration of the SDE is explained in \cite{BrownJr1963ThermalParticleb,evans2014atomistic,berkov2007magnetization}.

The time-dependent response of the MFTJ is calculated using the following procedure. \textit{\textbf{Step 1:}} the term $\frac{dF}{dP}$ is calculated from the LD equation, given the initial polarization $P$ and the external applied voltage $V_a$. The $\frac{dF}{dP}$ obtained in the previous step is substituted in the LK equation to get $\frac{dP}{dt}$. Then the forward difference formula is used to update the polarization as following $P(t+dt)=P(t)+\frac{dP}{dt}dt$. \textit{\textbf{Step 2:}} the electrostatic potential is calculated from (1) and (4). \textit{\textbf{Step 3:}} The spin current is calculated from the quantum transport model.  The magnetization obtained from the solution of LLG and the electrostatic potential obtained in the previous step are used to solve the quantum transport. \textit{\textbf{Step 4:}} the exchange coefficient $J'_{exch,i,j}$ is calculated from (\ref{eqn_a_t})-(\ref{eqn_Jprim}) using the electric polarization at time $t$ obtained from the LK equation.  \textit{\textbf{Step 5:}} the LLG equation is solved to get $m(t+dt)$ using the spin current obtained from the quantum transport. Finally, the steps (Step 1) to (Step 5) are repeated iteratively at each time step.

\jpapT{The transport parameters are usually estimated by fitting the parameters on the experimental I-V characteristic \cite{quindeau2015origin,gruverman2009tunneling,Datta2012Voltage}. However, the estimation process is not straight forward and more than one solution could produce the same transport properties. In this study, we try to go beyond that and predict some of the parameters from DFT calculations to reduce the complexity of the estimation process. On the other hand, some transport processes cannot be included in the DFT calculation. Note, the DFT by definition describes the system at the ground minimum energy state. In contrast, the quantum transport model exhibits nonequilibrium conditions by definition. Therefore, it is better to estimate certain parameters from experimental data to account for these limitations of DFT. Because of the aforementioned discussion, we adopted a combination of estimating the parameters directly from experimental data and estimating the parameters from DFT to improve the quality of parameter estimation.}

The simulation parameters used in this study are selected according to the following criterion. The saturation magnetization, magnetic exchange constant are estimated from DFT calculations. The magnetic anisotropy is selected from the experimental study \cite{Yin2015}. The magnetic damping factor is set within the acceptable range of similar structure. The screening lengths and the splitting energy are estimated by the DFT calculation. The effective mass are tuned within the acceptable range in literature to produce the experimental results. This is a very common procedure for selecting effective mass \cite{quindeau2015origin,gruverman2009tunneling,Datta2012Voltage, Datta2010QuantitativeDevices}.

\jpapT{The Landau-Devonshire equation parameters $\alpha_{1}$, $\alpha_{11}$, and $\alpha_{111}$ are calculated from the critical voltage at which the electric polarization is switched and the values of the polarization at zero voltage. The values of the critical voltage and polarization at zero voltage are known from the experimental results in \cite{Yin2015}. The two values of the polarization at zero voltage are local minimum points of the free energy. The free energy has a maximum point at zero polarization. The maximum and minimum points are located at the zeros of the first derivative of the free energy and impose constraints on the sign of the second derivative of the free energy. In addition, the coefficient of the highest order term in the free energy has to be positive because the free energy has to reach positive infinity as the polarization reaches $\pm\infty$. We used a numerical grid search to solve for $\alpha_{1}$, $\alpha_{11}$, and $\alpha_{111}$ that considers the aforementioned constraints.} The viscosity coefficient is estimated by a grid search to get $5ns$ switching time. Finally, the energy $E_a$ and $E_b$ in (\ref{eqn_a_t}) are estimated from DFT calculations.

\section{Simulation Results and Analysis}\label{ch_results}

\subsection{Comparison with Experimental MFTJ Characteristics}

\begin{figure}[!ht]
\centering
\includegraphics[width=0.8\linewidth,trim={0cm 0cm 0cm 0cm},clip=true]{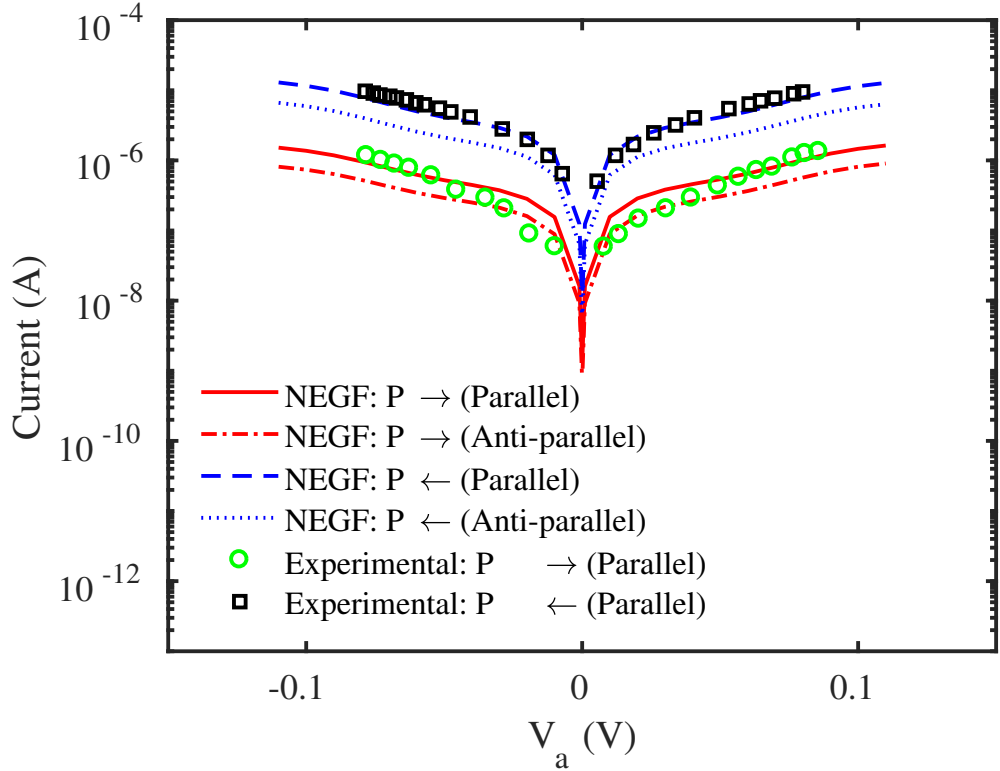}
\caption{The current of LSMO/LCMO/BTO/LSMO MFTJ is illustrated at different applied bias voltages $V_a$ for the four MFTJ states. The experimental data for the same device is demonstrated in \cite{Yin2015}.  The simulation parameters are $m_{ch}^*=0.8m_0$, $m_{L}^*=0.9m_0$, $m_{R}^*=0.9m_0$, $m_0$ is the free electron mass, $\mu_L=3$ $eV$, $\mu_R=3$ $eV$ $\Delta_L=2.4$, $\Delta_R=2.4$ $eV$, $U_B=3.1$ $eV$, $\phi_{BI}=1$ $eV$ $t_{FE}=2$ $nm$, $T=80 K$ \cite{Yin2015}, and the MFTJ radius is $8.5$ $\mu m$. The screening lengths of the electrodes used in NEGF simulation are estimated by DFT, as illustrated in section \ref{ch_dft}.}
\label{figure1_fit_LCMO_BTO_LSMO}
\end{figure}

\begin{figure}[!ht]
\centering
\includegraphics[width=0.8\linewidth,trim={0cm 0cm 0cm 0cm},clip]{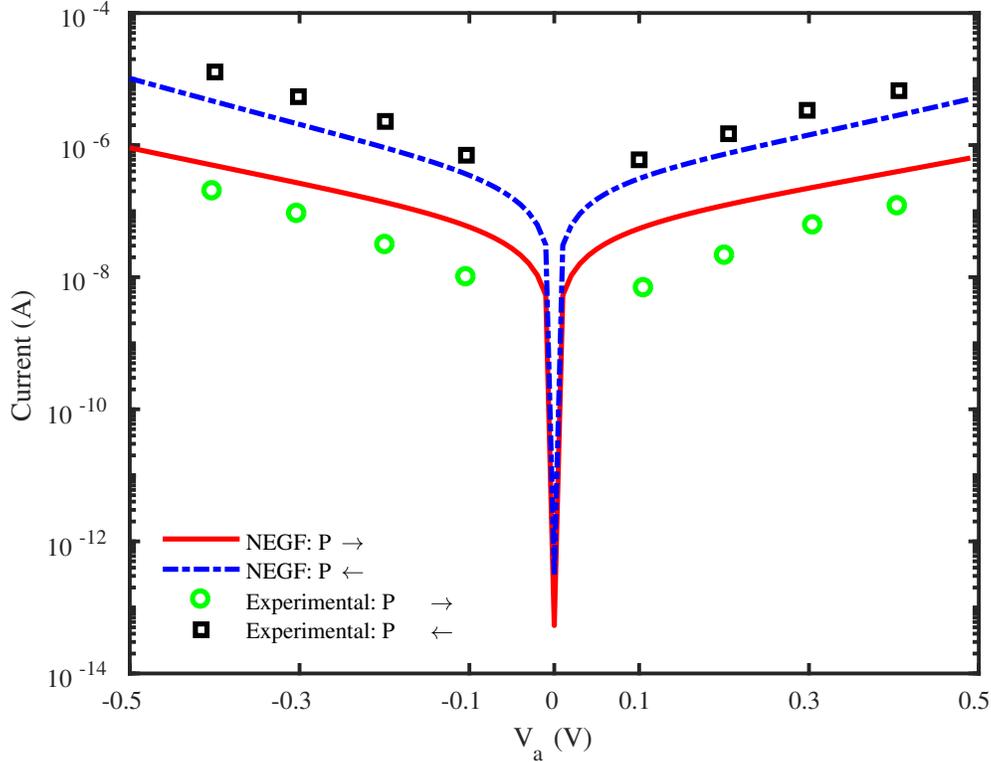}

\caption{The current of LSMO/BTO/Co FTJ is illustrated at different applied bias voltages $V_a$. The experimental data for the same device is demonstrated in \cite{Chanthbouala2012Solid-state}. The simulation parameters are $m_{ch}^*=0.8m_0$, $m_{L}^*=0.9m_0$, $m_{R}^*=2m_0$, $\mu_L=3$ $eV$, $\mu_R=2.9$ $eV$, $\Delta_L=2.4$, $\Delta_R=1.8$ $eV$, $U_B=3.16$ $eV$, $\phi_{BI}=1.2$ $eV$ $t_{FE}=3$ $nm$, $m_0$ is the free electron mass, $T=300 K$ \cite{Chang2017}, and the FTJ radius is $350$ $nm$. The screening lengths of the electrodes used in NEGF simulation are estimated by DFT, as illustrated in section \ref{ch_dft}.}
\label{figure0_fit_LSMO_BTO_Co}
\end{figure}

The main advantage of spin-based NEGF is the ability to model the four resistance states of the MFTJ. The LSMO/LCMO/BTO/LSMO MFTJ in \cite{Yin2015} is the most appropriate device to demonstrate these physical characteristics. The simulation results of LSMO/LCMO/BTO/LSMO are illustrated in Fig. \ref{figure1_fit_LCMO_BTO_LSMO}. The proposed framework can capture the majority of the MFTJ I-V characteristic for positive (low resistance) and negative (high resistance) polarization states, as illustrated in Fig. \ref{figure1_fit_LCMO_BTO_LSMO}. The simulation results for LSMO/BTO/Co FTJ are in agreement with the experimental results \cite{Chanthbouala2012Solid-state}, as illustrated in Fig. \ref{figure0_fit_LSMO_BTO_Co}. The simulation parameters of the LK and the LLG equations are $\alpha_1=-1.0654\times 10^9$ $m/F$,  $\alpha_{11}=-6.0878\times 10^9$ $m^5/(C^2.F)$, $\alpha_{111}=5.0499 \times 10^{10}$  $m^9/(C^4.F)$, $dt=1\times10^{-14}$ s, $K_U=7.8 \times 10^4$ $erg/cm^3$, $\alpha=0.05$, $\lambda=1.8$ $s/F$, $j_{exch,i,j}= 4.5\times 10^5$ $erg/cm^3$, $|E_a-E_b|=0.029$ $eV$ and $M_s=414.15$ $emu/cm^3$.

\begin{figure}[!h]
\centering
\includegraphics[width=0.8\linewidth,trim={0cm 0cm 0cm 0cm},clip=true]{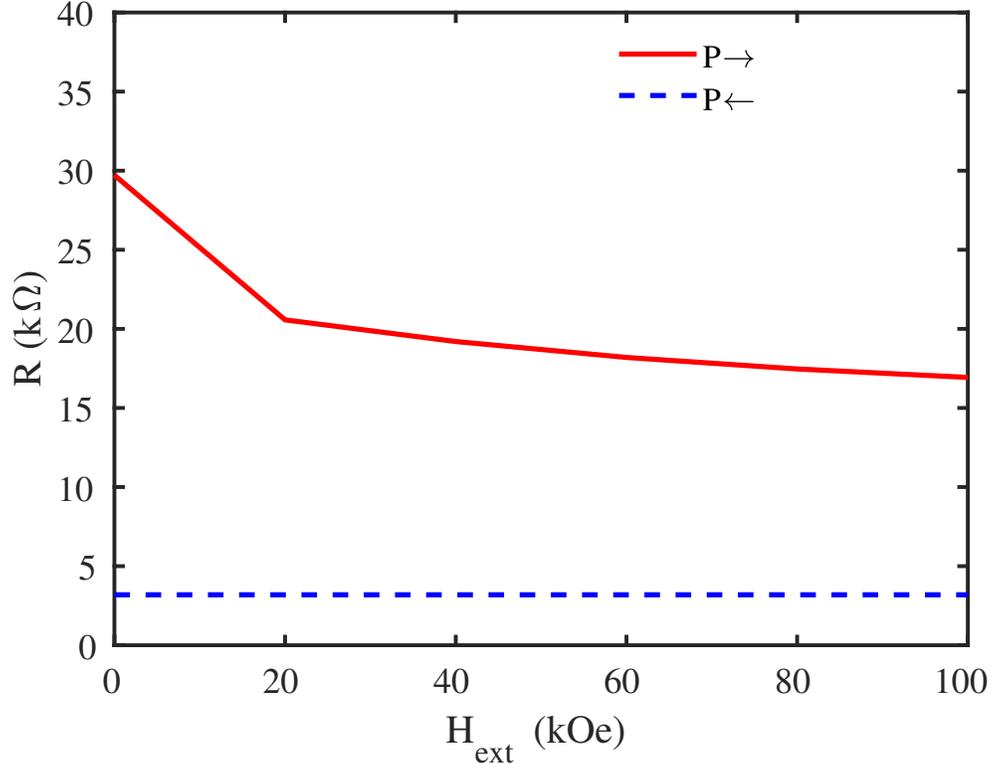}
\caption{The resistance of LSMO/LCMO/BTO/LSMO MFTJ is illustrated under the effect of external magnetic field.}
\label{figure4_R_H}
\end{figure}

As explained in section \ref{ch_dft}, the LCMO layer goes through a phase transition from the FM to the AFM alignment by the influence of the electric polarization switching. \ze{The phase transition is confirmed by the following experimental procedure \cite{Yin2015}.} Starting by applying an external magnetic field to the MFTJ, \sework{the} change in the resistance of the MFTJ due to the increase of the external magnetic field is measured. In the case of positive polarization, the device resistance diminishes due to the increase of the external magnetic field. In contrast, the negative polarization state exhibits a constant resistance independent of the magnetic field. This behavior of the MFTJ resistance is explained by the influence of the external magnetic field on the AFM aligned Mn site and the ability of the external magnetic field to switch the AFM aligned Mn site back to FM alignment \cite{Yin2015}. Fig. \ref{figure4_R_H} illustrates the effect of the external magnetic field on the MFTJ resistance that is produced by the \sework{quantum transport and magnetization dynamics}. The simulation mimics the same physical device behavior because the FM electrode Hamiltonian has a magnetization dependent term (\ref{eqn_h_LR}).

\subsection{Analysis of Various MFTJ Parameters}

\begin{figure}[!h]
\centering
\includegraphics[width=0.8\linewidth,trim={0cm 0cm 0cm 0cm},clip]{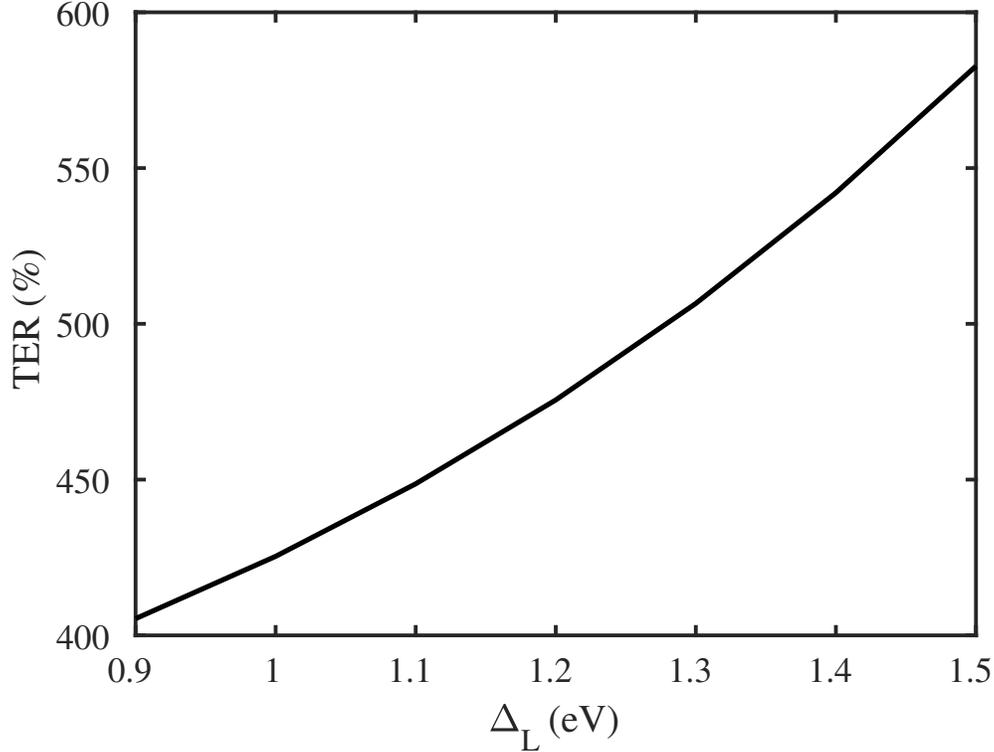}
\caption{The TER of the MFTJ as a function of splitting energy $\Delta_L$.}
\label{figure9_TER_TMR_delta}
\end{figure}

The TER is estimated at different values of the splitting energy, as illustrated in Fig. \ref{figure9_TER_TMR_delta}. The TER dependence on splitting energy originates from the ME effect that happens in the LCMO/BTO interface. Moreover, the $TMR_\rightarrow$ is consistently lower than $TMR_\leftarrow$ as illustrated in Fig. \ref{figure1_fit_LCMO_BTO_LSMO}. This asymmetric behavior is due to the antiferromagnetic alignment of the Mn sites of the LCMO electrode in the case of $P_\rightarrow$ state that reduces the TMR effect at that state. In contrast, the LCMO exhibits FM alignment in the case of $P_\leftarrow$ state, and hence the $TMR_\leftarrow$ is higher compared to $TMR_\rightarrow$.

\begin{figure}[!h]
\centering
\includegraphics[width=0.8\linewidth,trim={0cm 0cm 0cm 0cm},clip=true]{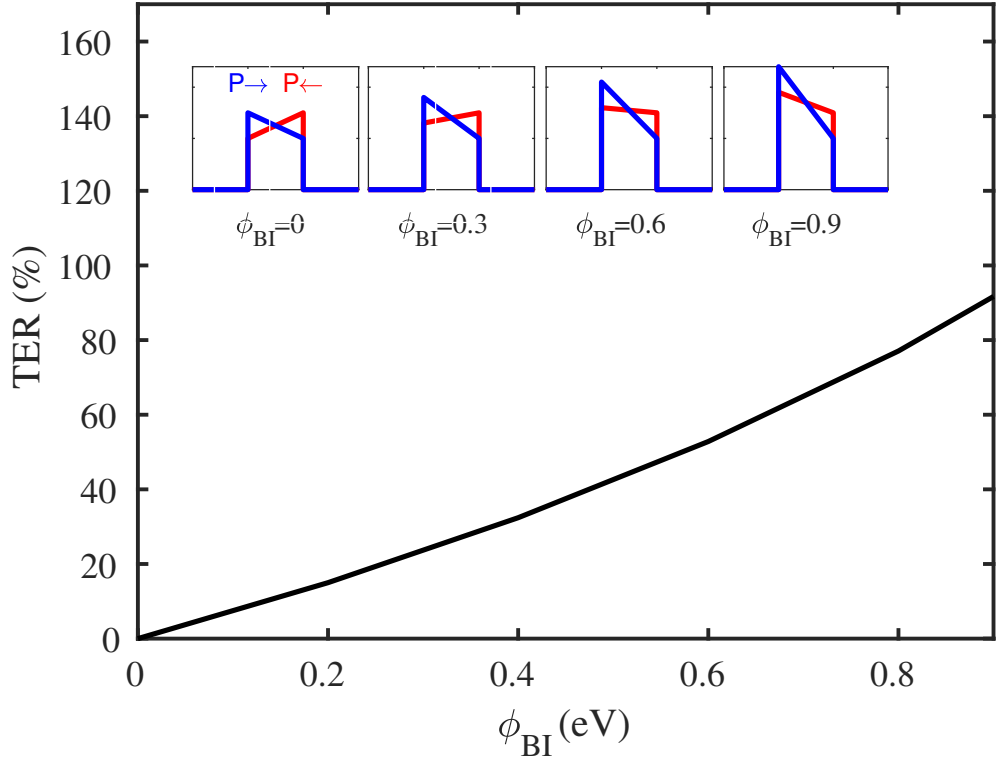}
\caption{The TER as a function of the built-in potential $\phi_{BI}$.}
\label{figure11_VBI}
\end{figure}

The asymmetry in the electrodes screening lengths is \sework{necessary} for an FTJ to exhibit a TER effect, as explained in section \ref{ch_formalism}. However, the NEGF \ze{transport} simulations show a significant TER ratio for a hypothetical device that has electrodes of identical screening lengths. Fig. \ref{figure11_VBI} shows the TER ratio as a function of the built-in potential $\phi_{BI}$ along with the electrostatic potential of the positive and negative polarization states at different values of $\phi_{BI}$. Although the electrodes screening lengths are identical, the TER ratio increases significantly due to the increase of the built-in potential. The origin of the TER effect in the case of symmetric electrodes screening lengths could be explained by observing the potential profile of the positive and negative polarization states. In the case of $\phi_{BI}=0$, the potential profiles of $P_\rightarrow$ and $P_\leftarrow$ are symmetric, and hence the TER ratio is zero as expected. However, as the built-in potential $\phi_{BI}$ increases, the potential profile of $P_\rightarrow$ and $P_\leftarrow$ start to deviate from the symmetric shapes to asymmetric potential profiles that have different average barrier height, as illustrated in the inset of Fig. \ref{figure11_VBI}. In other words, the built-in potential introduces another source of asymmetry that allows the modulation of the barrier average height by the electric polarization. 

\begin{figure}[!h]
\centering
\includegraphics[width=0.8\linewidth,trim={0cm 0cm 0cm 0cm},clip]{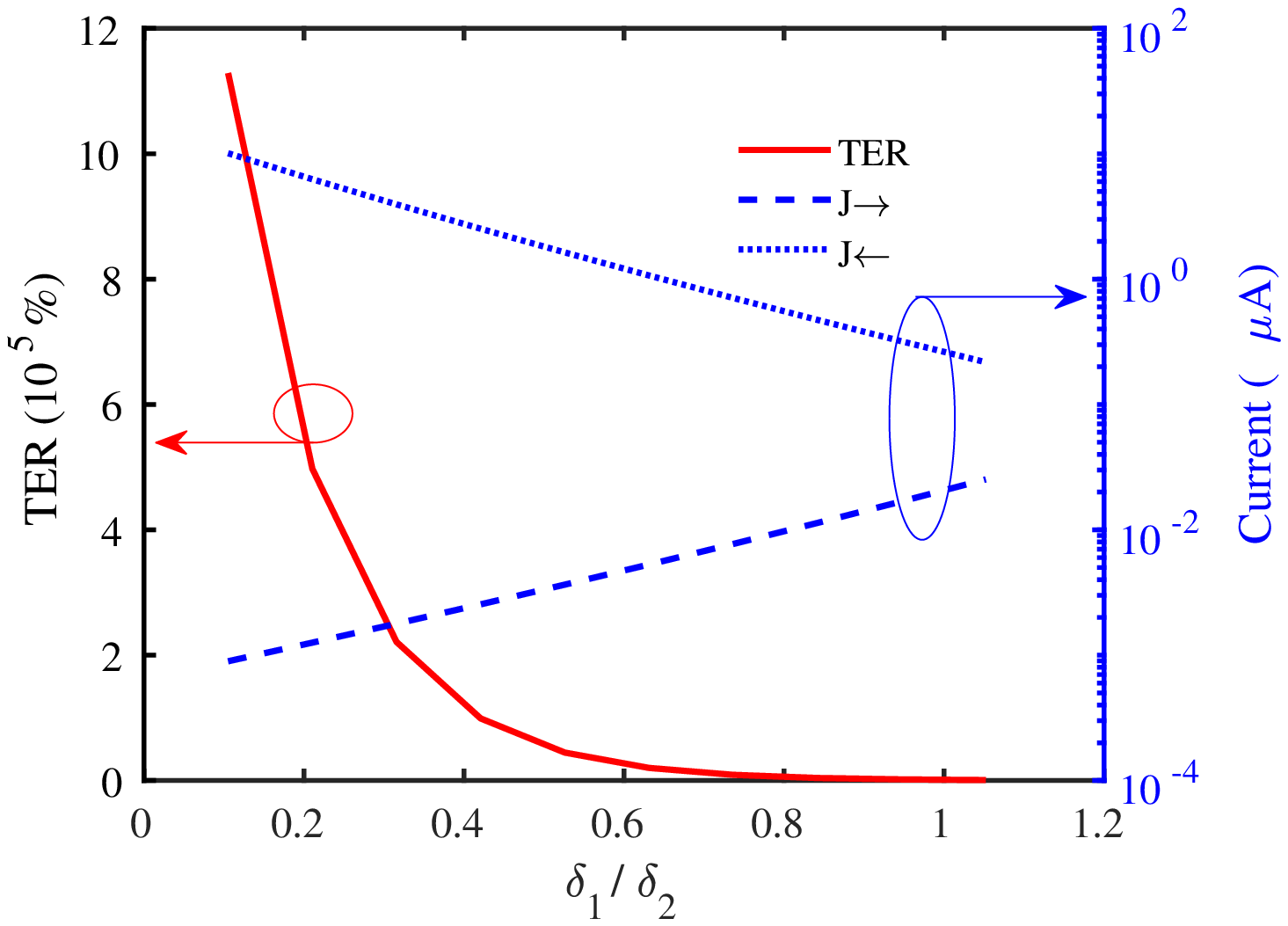}
\caption{The TER of the MFTJ as a function of the screening lengths ratio $\delta_1/\delta_2$.}
\label{figure10_TER_TMR_epr1}
\end{figure}

The exponential dependence of the TER on the left and right electrode screening lengths ratio is demonstrated in Fig. \ref{figure10_TER_TMR_epr1}. \sework{In order to understand the TER behavior as a function of $\frac{\delta1}{\delta2}$, we have \sework{reordered} the TER definition as $TER=\frac{j_\leftarrow}{j_\rightarrow}-1$, where $j_\rightarrow$ and $j_\leftarrow$ are the current at positive (\textit{high resistance}) and negative (\textit{low resistance}) polarization states, respectively. The strength of the barrier height modulation, induced by polarization switching, is enhanced by increasing the difference between the screening lengths of the electrodes. Therefore, the current $j_\rightarrow$ increases and the current $j_\leftarrow$ decreases, as illustrated in  Fig. \ref{figure10_TER_TMR_epr1}. Consequently, the TER improves exponentially as the ratio $\frac{\delta1}{\delta2}$ reaches zero.} The same conclusion can be quantitatively \sework{derived} from the Thomas-Fermi relation (\ref{eqn_thomas_fermi}) that formulates the potential at the interface as $\phi_1=\frac{\sigma_s \delta_1}{\epsilon_0\epsilon_{r1}}$. Therefore, the potential $\phi_1$ decreases as $\delta_1$ shrinks, and hence the the potential difference $||\phi_1|-|\phi_2||$ rises along with the \sework{ratio $\frac{j_\leftarrow}{j_\rightarrow}$. As a result of $\frac{j_\leftarrow}{j_\rightarrow}$ exponential increase, the TER exponentially improves}.

\begin{figure}[!ht]
\centering
\includegraphics[width=0.8\linewidth,trim={0cm 0cm 0cm 0cm},clip]{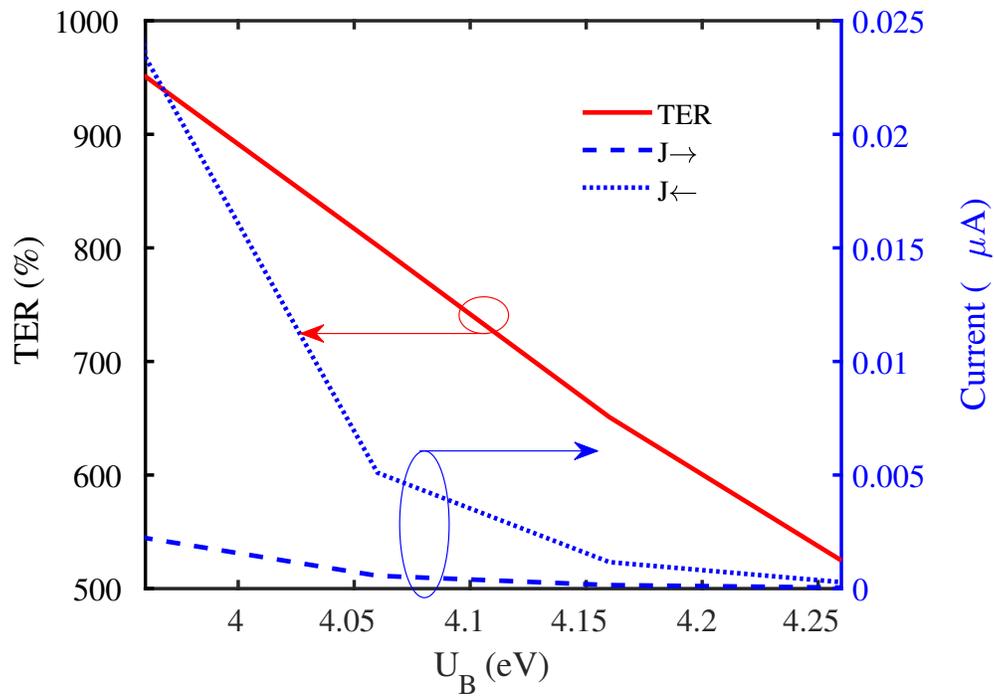}
\caption{The current and TER of the MFTJ as a function of barrier height $U_B$.}
\label{figure6_TER_Ub}
\end{figure}

Interestingly, the TER shows exponential dependence on the ratio $\frac{\delta_1}{\delta_2}$, but weaker dependence on the barrier height $U_B$. \ze{The rationale behind the difference in the dependence of TER on $\frac{\delta_1}{\delta_2}$ and $U_B$ is explained in the following comparative analysis.} The decay of the ratio $\delta_1/\delta_2$ results in increasing  $j_\rightarrow$ and decreasing $j_\leftarrow$ that exponentially enhance the TER. In contrast, the increase in the barrier height $U_B$ reduces both $j_\rightarrow$ and $j_\leftarrow$ but \ze{with} different rates. \ze{Therefore, the TER changes with a weak rate because it is proportional to $\frac{j_\leftarrow}{j_\rightarrow}$, as illustrated in Fig. \ref{figure6_TER_Ub}.} However, the TER curve looks approximately linear because of the narrow range of $U_B$ along with the weak exponential dependence of the TER on $U_B$. The MFTJ high and low resistances is exponentially augmented as $U_B$ elevates, as observed from the currents $j_{\rightarrow}$ and $j_{\leftarrow}$ in Fig. \ref{figure6_TER_Ub}. The barrier height is dependent on the insulator and electrodes work functions. \ze{Therefore, the electrodes work functions together with the screening lengths have a strong influence on the resistance and the TER of MFTJ, respectively.}

\subsection{MFTJ Dynamic Characteristics}
\begin{figure}[!h]
\centering
\includegraphics[width=1\linewidth,trim={2.5cm 8.5cm 2.5cm 8.5cm},clip]{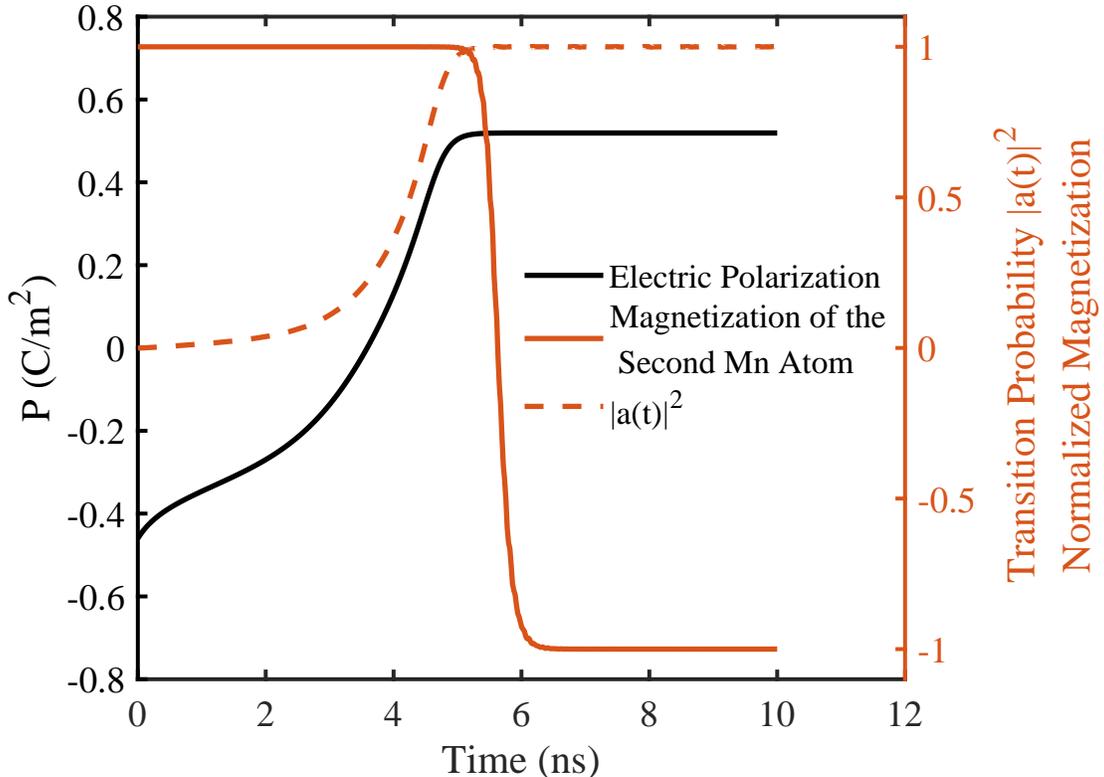}
\vspace{-2\baselineskip}
\caption{The electric polarization, the transition probability $|a(t)|^2$ and the $z$ component of the magnetization of the second Mn atom away from the LAMO/BTO interface.}
\label{fig_Trans_Prob}
\end{figure}

\begin{figure}[!ht]
\centering
\includegraphics[width=1\linewidth,trim={2.5cm 8.5cm 2.5cm 8.5cm},clip]{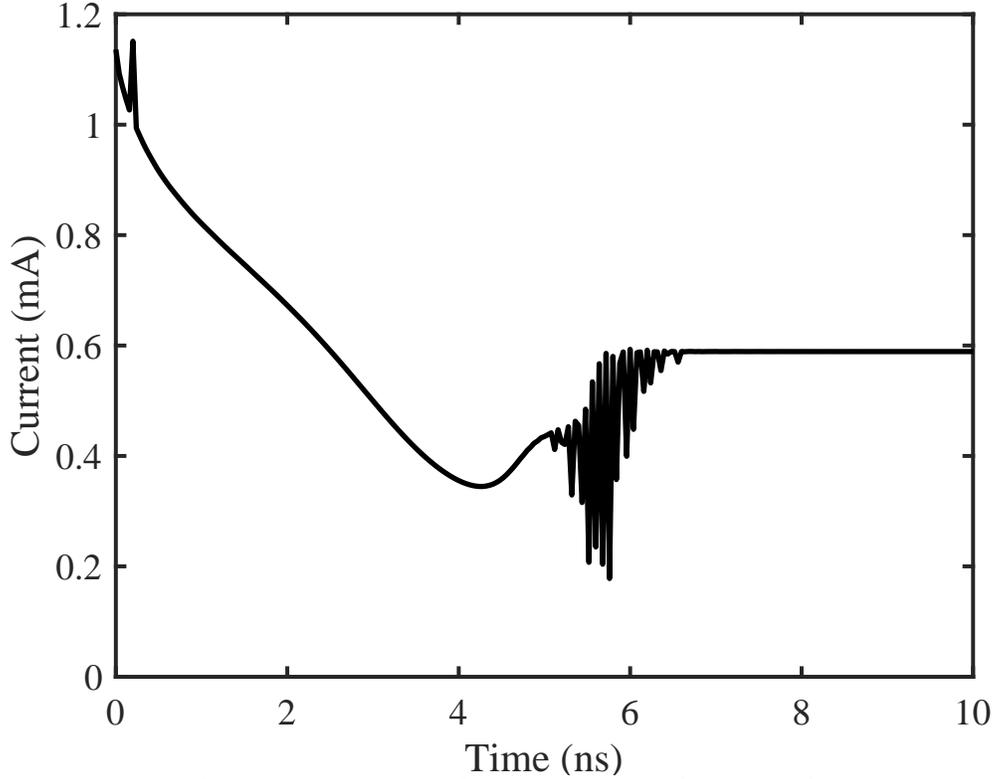}
\vspace{-2\baselineskip}
\caption{The current of the MFTJ as a function of time.}
\label{fig_Trans_current}
\end{figure}

The time-dependent response of the MFTJ switching is illustrated in Fig. \ref{fig_Trans_Prob}. The MFTJ dynamic response is calculated according to the procedure explained in section \ref{ch_procedure}. The electric polarization takes around $5ns$ to switch from from negative to positive value and saturate.  The transition probability elevates from zero to one and saturates following the electric polarization, as illustrated in Fig. \ref{fig_Trans_Prob}. The transition probability of the FM phase to the AFM phase is calculated from (\ref{eqn_a_t}) that is based on the time-dependent perturbation theory. The exchange coupling coefficient follows the transition probability according to (\ref{eqn_Jprim}). Therefore, the exchange coefficient change from a positive value (FM alignment) to a negative value (AFM alignment). The magnetization of the second Mn atom switches from positive to negative magnetization following a magnetic precession motion because of the change in the exchange coupling. The precession motion of the magnetization causes the oscillation of the MFTJ current, as illustrated in Fig. \ref{fig_Trans_current}.

Interestingly, the electric current decreases significantly after $4ns$ from the start of the switching process. The large variations of the current are due to the electric polarization switching. In contrast, the change in magnetic configuration lags the electric polarization switching, as illustrated in Fig. \ref{fig_Trans_Prob} and Fig. \ref{fig_Trans_current}.  During the switching process, the electric polarization changes from negative to positive direction passing by $P=0$. The electrostatic potential is modulated by the electric polarization as described by Thomas-fermi relation. The current reaches its minimum value because the tunneling current changes from Nordheim tunneling to direct tunneling at that point. The current changes back to Nordheim tunneling after the minimum point. Therefore, the current starts to increase after the minimum current point.  Note, non-ideal contacts are assumed at high switching voltage to allow a constant voltage drop at each contact of $0.2V$ during  quantum transport.

In contrast, the oscillations that start after the current minimum point are due to the precession motion of the Mn atoms.  The precession motion of the Mn atoms is derived by the magnetic exchange torque of neighbor atoms. Due to the time-dependent perturbation potential caused by electric polarization switching, the probability of magnetization switching to AFM alignment increases from zero to one. The magnetization switching lags the switching probability by hundreds of picoseconds. The reason behind the delay in the magnetization switching is that the magnetization and the effective exchange field have almost an angle of $\pi$ at the initial position. Therefore, the magnetization motion under the effect of magnetic torque is slow at the beginning. The thermal fluctuations could assist the switching process at the slow-starting part of the switching process. However, the thermal fluctuations are small compared to the magnetic anisotropy due to the large area of the MFTJ.

\section{Conclusion} \label{conc}

In this study, we propose a modeling and simulation framework that captures the behavior of MFTJ as a four-state device. Furthermore, the DFT method is used to estimate the screening length of the electrodes that have a strong influence on the TER. The estimated screening lengths are used in the quantum transport calculations to mimic the realistic device behavior at different bias voltages. The TER and TMR estimated by the proposed framework is compared with the experimental results of LSMO/LCMO/BTO/LSMO \cite{Yin2015}. The quantum transport and magnetization dynamics could show the dependence of the device resistance on the external applied magnetic field. The dependence of the MFTJ resistance on the external magnetic field is in agreement with the experimental results in \cite{Yin2015} that confirms the transition from the FM to the AFM phase in the LCMO electrode.

Our analysis illustrates that not only the TMR but also the TER of the MFTJ depends on splitting energy because of the magnetoelectric effect at the interface that originates from the LCMO electrode phase transition from the FM to the AFM phase. On the other hand, the contrast between the weak and strong exponential dependence of the TER on the barrier height and electrodes screening length ratio, respectively is analyzed. The barrier height that is dependent on the electrodes and insulator work functions, could increase the MFTJ high and low resistance. Consequently, the power and speed of the MFTJ sensing could be enhanced by choosing the insulator and electrodes that have the appropriate work functions. However, the ratio of the electrodes screening length $\delta_1/\delta_2$ could exponentially enhance the TER ratio. Finally, our analysis reveals that the TER effect is improved by the asymmetry exhibited by the built-in potential that results in average barrier height modulation by electric polarization.

Based on the time-dependent perturbation theory, we could derive a mathematical formulation that relates the magnetic exchange interaction coefficient to the time evolution of electric polarization  (\ref{eqn_a_t})-(\ref{eqn_Jprim}). \jpapT{This formulation is an important step toward a consistent model of MFTJs. The formulation emphasizes that the magnetization switching from FM to AFM alignment induced by polarization reversal follows a precessional motion.  The transient response of the MFTJ exhibits a transition from Nordheim tunneling to direct tunneling and back to Nordheim tunneling current during the polarization switching. The transient response of the MFTJ demonstrates a set of oscillations due to the magnetic precessional motion. Although the switching from the FM to AFM alignment is induced by the electric polarization switching, the thermal magnetic fluctuations still assist the magnetization motion especially at the start of the switching.} 

\section*{Acknowledgement}
The work was supported in part by Semiconductor Research Corporation(SRC), the National Science Foundation, Intel Corporation and by the DoD Vannevar Bush Fellowship.

\bibliography{MyCollection3.bib}
\end{document}